\documentclass[10pt, conference, letterpaper]{IEEEtran}

\usepackage{ifpdf}
\usepackage{cite}
\usepackage{algorithm}
\usepackage{algorithmic}
\usepackage{subfigure}
\usepackage{amsmath}
\usepackage{algorithmic}
\usepackage{array}
\usepackage{stfloats}
\usepackage{url}
\usepackage{color}

\usepackage{authblk}

\ifCLASSINFOpdf
  \usepackage[pdftex]{graphicx}
\else
\fi

\IEEEoverridecommandlockouts\IEEEpubid{\makebox[\columnwidth]{979-8-3503-9973-8/23/\$31.00 $\copyright$2023 IEEE\hfill}\hspace{\columnsep}\makebox[\columnwidth]{ }}  

\begin{document}

\title{Boosting Distributed Machine Learning Training Through Loss-tolerant Transmission Protocol
\thanks{* Corresponding author: Yang Xu (xuy@fudan.edu.cn)}
\thanks{First author: Zixuan Chen (zxchen20@fudan.edu.cn)}
\thanks{This work is sponsored by the Key-Area Research and Development Program of Guangdong Province (2021B0101400001), National Natural Science Foundation of China (62150610497, 62172108, 62002066), Natural Science Foundation of Shanghai (23ZR1404900), Shanghai Pujiang Program (2020PJD005), the Major Key Project of PCL (PCL2021A15), and Open Research Projects of Zhejiang Lab (2022QA0AB07).
}
\thanks{This paper will be published on IWQoS 2023. Preview version only.}
}

\author[$\dagger$]{Zixuan Chen}
\author[$\dagger$]{Lei Shi}
\author[$\dagger$]{Xuandong Liu}
\author[$\dagger$]{Xin Ai}
\author[$\dagger \ddagger$]{Sen Liu}
\author[$\dagger \diamondsuit \ast$]{Yang Xu}
\affil[$\dagger$]{School of Computer Science, Fudan University, Shanghai, China}
\affil[$\ddagger$]{Institute of Fintech, Fudan University, Shanghai, China}
\affil[$\diamondsuit$]{Peng Cheng Laboratory, Shenzhen, China}

\maketitle

\begin{abstract}

Distributed Machine Learning (DML) systems are utilized to enhance the speed of model training in data centers (DCs) and edge nodes. The Parameter Server (PS) communication architecture is commonly employed, but it faces severe long-tail latency caused by many-to-one "incast" traffic patterns, negatively impacting training throughput. To address this challenge, we design the \textbf{L}oss-tolerant \textbf{T}ransmission \textbf{P}rotocol (LTP), which permits partial loss of gradients during synchronization to avoid unneeded retransmission and contributes to faster synchronization per iteration. LTP implements loss-tolerant transmission through \textit{out-of-order transmission} and \textit{out-of-order Acknowledges (ACKs)}. LTP employs \textit{Early Close} to adjust the loss-tolerant threshold based on network conditions and \textit{bubble-filling} for data correction to maintain training accuracy. LTP is implemented by C++ and integrated into PyTorch. Evaluations on a testbed of 8 worker nodes and one PS node demonstrate that LTP can significantly improve DML training task throughput by up to 30x compared to traditional TCP congestion controls, with no sacrifice to final accuracy.

\end{abstract}


%
\IEEEpeerreviewmaketitle

\section{Introduction}

With the explosion of dataset and model size in Machine Learning (ML) applications, Distributed Machine Learning (DML) has been widely adopted to leverage the power of multiple worker nodes during large-scale training. To achieve synchronization among the distributed worker nodes, several DML communication architectures, such as the Parameter Server architecture (PS)~\cite{li2013parameter} and Ring-AllReduce~\cite{gibiansky2017bringing}, are proposed, improving the efficiency of the distributed training systems.

\begin{figure}[tbp]
\centerline{\includegraphics[width=9cm]{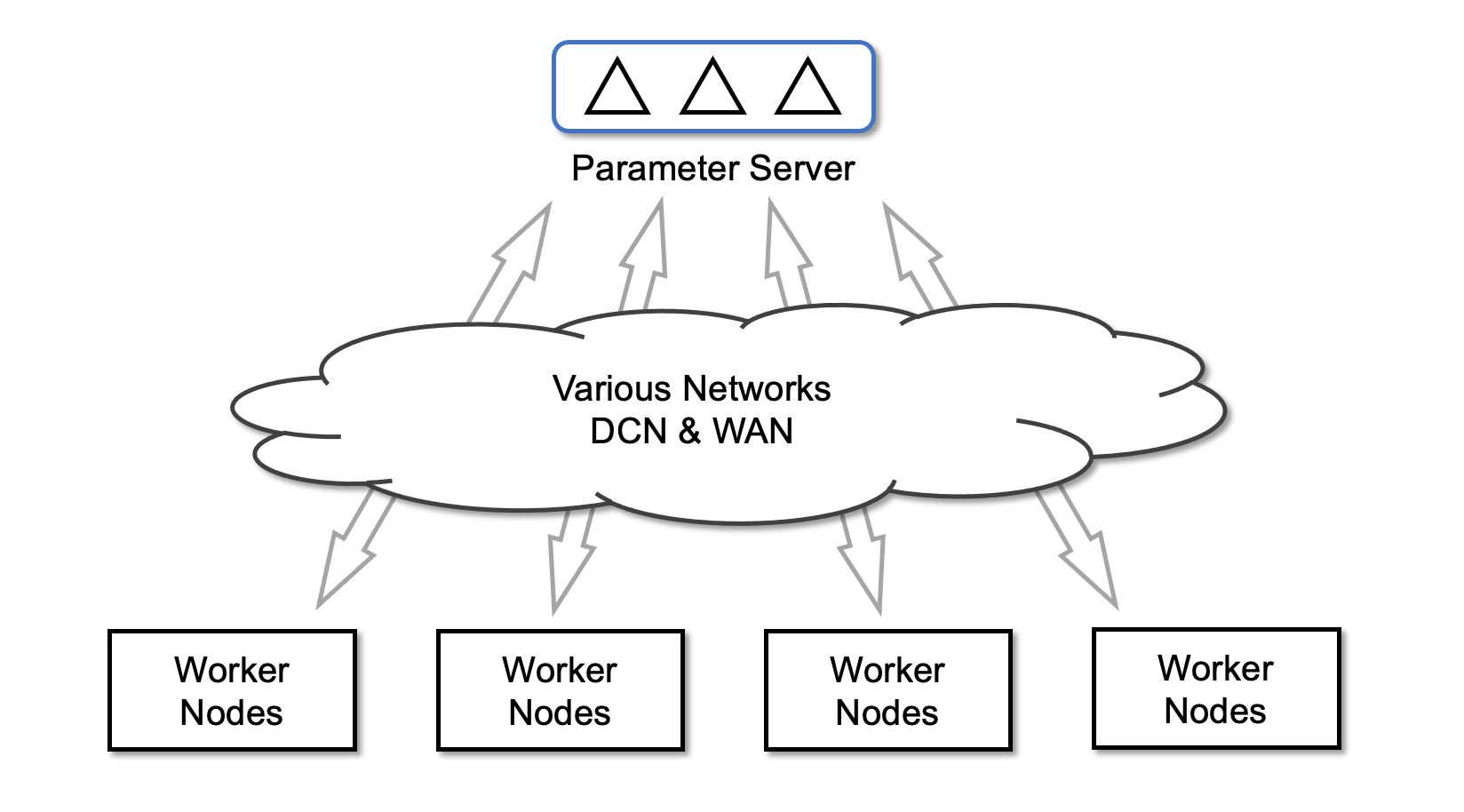}}
\caption{PS architectures in DML training.}
\vspace{-0.2in}
\label{fig:multi-links-on-PS}
\end{figure}

The PS architecture has become a prevalent communication architecture in DML due to its simplicity and efficiency. In the PS architecture, all computing nodes, also known as worker nodes, are managed by one or multiple PS(es). The worker nodes learn a portion of the training dataset and communicate with the associated PS to synchronize training results with other worker nodes (as shown in Figure~\ref{fig:multi-links-on-PS}). During each iteration, each worker node trains the model using a portion of the dataset and sends the computed gradients to the PS, which aggregates the gradients from all worker nodes and updates the global model. Finally, the PS sends the latest model to each worker node, preparing it for the next iteration. The parallel computation of multiple worker nodes greatly improves the efficiency of the model training process.

As the scale of the DML system increases, the network connecting the worker nodes and the PS is becoming a major limitation for the system. This is due to two main reasons. Firstly, most of the existing DML training tasks use the Bulk Synchronous Parallel (BSP) synchronization model~\cite{gerbessiotis1994direct}, in which all nodes must fully synchronize their gradients with the PS at the end of each training batch. This many-to-one incast traffic pattern results in long-tail latency problems~\cite{zats2012detail}, significantly reducing the communication time. Secondly, recently, several new DML-based scenarios have been proposed, such as ML across data centers~\cite{hsieh2017gaia}, edge computing~\cite{shi2016edge, lyu2019optimal}, and federated learning (FL)~\cite{mcmahan2017communication}. These scenarios typically involve transmitting gradients over wide area networks (WANs) or wireless networks, which pose new challenges of unstable links. As a result, non-congestion packet loss is common in these scenarios. For instance, micro-burst traffic in data center networks~\cite{shan2019observing, shan2018micro,zhang2020vms}, physical link failures (e.g., optical fiber)~\cite{li1997automatic, choi2002double}, wireless links at edge nodes~\cite{abbas2017mobile, guduz2019machine}, and re-routing in WANs~\cite{lor2010packet} can result in non-congestion packet loss and reduce the efficiency of communication between the PS and worker nodes. The long-tail latency generated by incast and non-congestion packet loss severly decreases the efficiency of DML training.

The current solutions to address the long-tailed latency issue caused by incast, such as pHost~\cite{gao2015phost} and Homa~\cite{montazeri2018homa}, do not cater to the specific requirements of DML training and are not effective in dealing with non-congestion packet loss. Moreover, DML allows for a certain level of data loss due to its numerical analysis process. To improve DML communication, several works have been proposed such as parameter quantization~\cite{lai2018enabling}, parameter pruning~\cite{yao2017deepiot}, gradient compression~\cite{aji2017sparse}, and gradient quantization~\cite{alistarh2017qsgd}. These methods aim to reduce the amount of data transferred per iteration by optimizing the communication process at the application level. However, these solutions still have limitations as they only reduce the communication size, but do not address the root cause of the long-tailed latency issue during synchronizations.

In this paper, we present a novel solution called the Loss-tolerant Transmission Protocol (LTP) aimed at enhancing the synchronization efficiency of DML training tasks. The protocol allows for partial data loss (loss-tolerant transmission) while synchronizing the gradients in the system, which can help alleviate the long-tail latency problems that commonly arise in DML systems. LTP enables loss-tolerant transmission through the use of out-of-order transmission and out-of-order ACK. To achieve this, LTP employs an Early Close mechanism to determine the threshold for loss-tolerant transmission and uses the bubble-filling mechanism to prevent data errors. To summarize, our contributions are

\begin{enumerate}
    \item We propose LTP, a transport protocol that improves the synchronization efficiency of DML training by allowing partial data loss. Two key mechanisms are designed, \textit{Early Close} and \textit{bubble-filling}. The \textit{Early Close} can finish the transmission earlier based on a pre-determined threshold on data percentage and transmission time (which we called loss-tolerant transmission), while the \textit{bubble-filling} preserves the correctness and accuracy of the machine learning tasks. A bandwidth-delay-product(BDP)-based congestion control algorithm ensures high bandwidth utilization in various network conditions.

    \item We implement LTP on Linux by C++ and integrate it into the widely used ML framework PyTorch. LTP is transparent to the ML framework, so DML programmers do not need to change their ML code. To use LTP as the communication protocol, the programmers only need to make simple modifications to the interfaces of the sockets.

    \item We evaluate the performance of LTP in various network conditions on a real testbed composed of 8 worker nodes and one PS node. Popular ML models (such as ResNet50 and VGG16) with the CIFAR10 dataset are used for evaluations. Evaluation results show that LTP can deliver up to 30x training speedup with no precision loss compared to conventional congestion control algorithms. 
\end{enumerate}

The rest of the paper is organized as follows. \S~\ref{sec:motivation} introduces the background and motivation of LTP. \S~\ref{sec:design} presents the design of LTP along with its two key mechanisms Bubble Filing and Early Close. \S~\ref{sec:implementation} presents the implementation of LTP and \S~\ref{sec:evaluation} presents the evaluation results. We conduct a discussion in \S~\ref{sec:discussion}. \S~\ref{sec:related_works} summarizes the related works and finally \S~\ref{sec:conclusion} concludes the paper.

\section{Background and Motivation}\label{sec:motivation}

\subsection{Network Becomes a Limitation to the Scaling of DML}

\begin{figure}[tbp]
\centerline{\includegraphics[width=8cm]{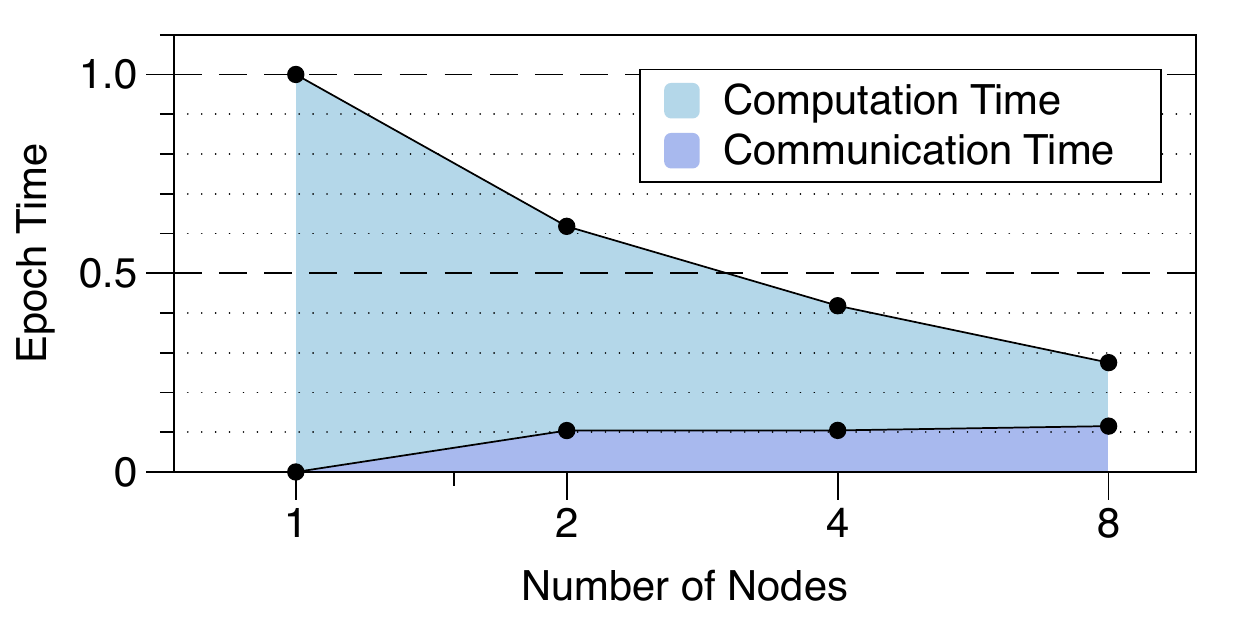}}
\caption{Scalability problem of DML training tasks: disproportionate reduction of training time and the number of worker nodes in DML training tasks.}
\label{fig:scale-of-dml}
\end{figure}

Although networks' bandwidth has increased rapidly in recent years, it is still a significant bottleneck for DML training. We evaluate the ResNet50~\cite{he2016deep} model for a DML training on 1, 2, 4, and 8 machines, respectively, with PS communication architecture. Figure~\ref{fig:scale-of-dml} shows that the training efficiency of DML does improve as the size of the computing nodes increases(the time in each epoch is decreasing), but the additional communication overhead also increases gradually(the ratio of communication time to computation time is increasing). The number of nodes is disproportionate to the optimization it brings to the whole training time.

We speculate that PS architecture's incast traffic causes this, i.e., multiple worker nodes communicate to only a few PSes in parallel and bursts. Although most nodes can finish their transmission almost simultaneously, some may suffer from the slow-growing rate of the congestion window (cwnd) because they are in a long-term competitive relationship. These lag flows will slow down the overall training synchronization, commonly found in DCNs and WANs.

\begin{figure}[htbp]
\centerline{\includegraphics[width=0.9\columnwidth]{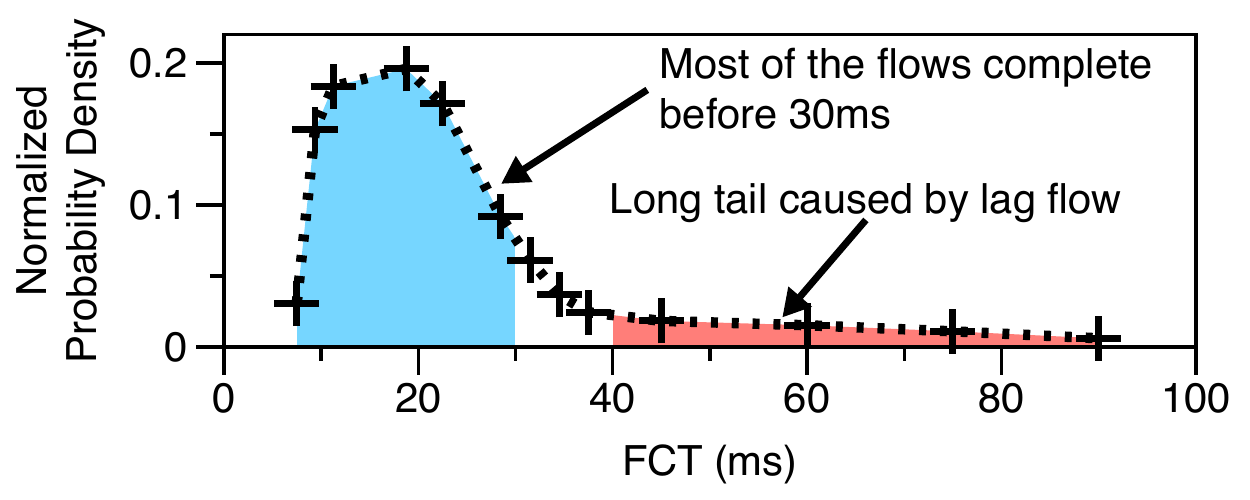}}
\caption{Long-tail latency caused by the incast traffic patterns.}
\label{fig:lag-flow}
\end{figure}

We do another experiment to illustrate the hazards of the long-tail latency. Under the default TCP protocol parameters, We use 8 worker nodes and one PS to build a many-to-one communication with a fixed message size and count the flow completion time of each worker node. Figure~\ref{fig:lag-flow} shows the probability density distribution of the FCTs(Flow Complete time). We can see that most of the flows have relatively similar FCT distributions, but there are still some ``starved" flows with relatively long FCTs. Since the existing popular DML training tasks still use the BSP synchronization model, the system will be blocked until all worker nodes complete the synchronization, slowing down the whole training throughput. 


\subsection{TCP Performs Poorly in Lossy Networks}\label{sec:tcp_poor_performance}

\begin{figure}[htbp]
\centerline{\includegraphics[width=9cm]{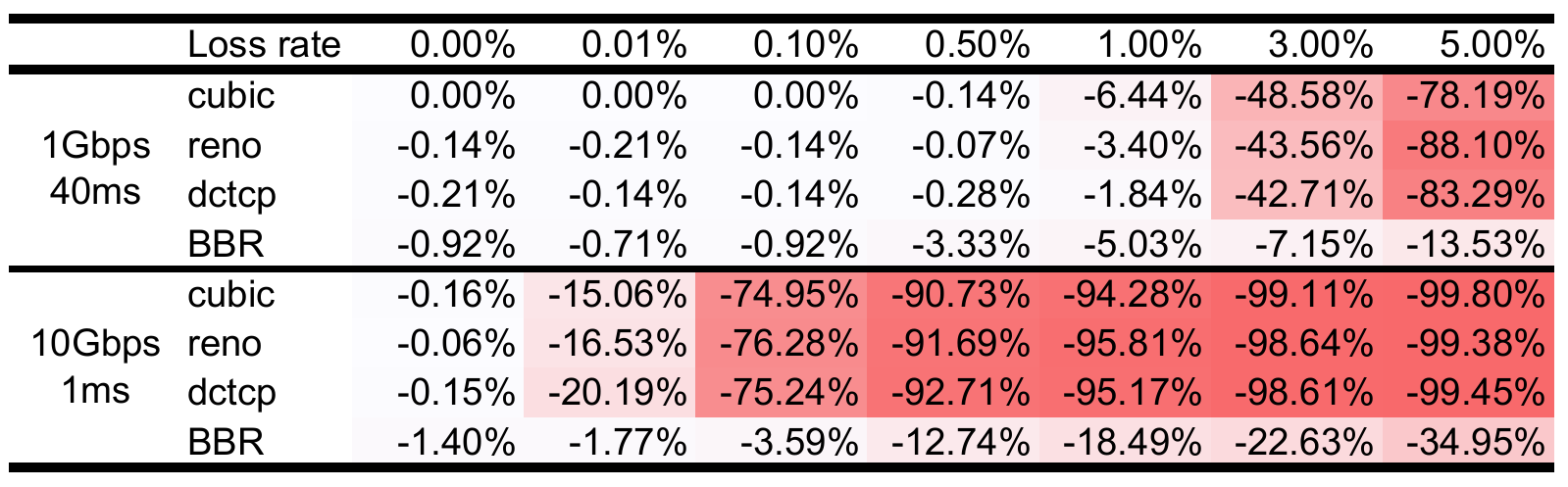}}
\caption{Different TCP congestion control in networks with non-congestion packet loss.}
\label{fig:different-tcp}
\end{figure}

After decades of development, the commonly used TCP congestion control algorithms have strong robustness. However, these congestion control algorithms have drawbacks in network environments where non-congestion packet loss exists. We conduct sets of experiments to evaluate the problems of TCP in unstable networks, which uses existing popular TCP congestion control algorithms to test the point-to-point pure traffic performance in an unstable network environment. We perform these experiments in both DCNs and WANs, respectively.

Figure~\ref{fig:different-tcp} is the bandwidth utilization reduction of different TCP congestion control algorithms in various networks, which shows that the conventional TCP congestion control algorithms perform poorly in the network with packet loss, especially in the network with higher bandwidth and lower latency. Although BBR\cite{cardwell2016bbr} can perform better in packet loss environments, the reduction is more significant than the non-congestion packet loss rate. Since traditional TCP congestion controls are order-preserving and use 3 duplicate ACKs as the signal that the queue of bottleneck link is full for congestion avoidance. LTP transmits the packets out of order and uses a BDP-based congestion control algorithm similar to BBR to maintain the link utilization, which has a better performance than the generally used congestion controls.

\subsection{Accuracy Degradation of Random Data Loss is Acceptable}\label{sec:selection-of-dropped-gradients}

DML is a numerical analysis process consisting of hundreds or thousands of iterations, so a certain threshold of data loss during each iteration will not affect the performance of the model. A number of methods have been proposed to accelerate the efficiency of DML training based on this loss tolerance property. They are divided into two main categories, including gradient quantization and sparsification~\cite{lin2017deep}. Gradient quantization approaches use low-bit floating numbers to store the data to be transmitted, e.g., a 32-bit floating number can be approximated by an 8-bit floating number, which can reduce its communication costs by a quarter. Gradient sparsification selects a portion of the data to transfer from the gradient vector. The Top-k~\cite{aji2017sparse} algorithm transmits only the absolute values of the first k large in the gradient vector. Random-k~\cite{stich2018sparsified} randomly transmits a part of the data from the gradient vector, which reduces the sorting overhead compared to Top-k. These two approaches can be combined. For example, DGC~\cite{lin2017deep} performs gradient sparsification with other training tricks like warm-up training~\cite{goyal2017accurate} and momentum correction~\cite{qian1999momentum}.

\begin{figure}[tbp]
\centerline{\includegraphics[width=8cm]{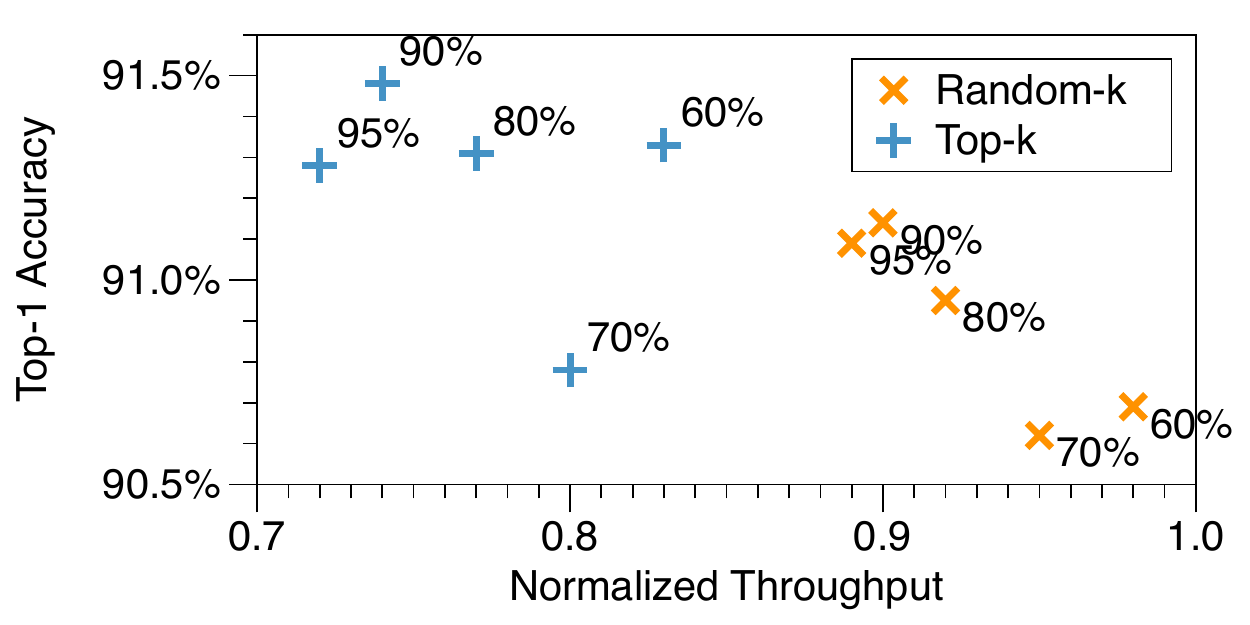}}
\caption{Comparison of top-1 accuracy and normalized throughput between Top-k and Random-k on ResNet18 with the CIFAR10 dataset.}
\label{fig:randk-and-topk}
\end{figure}

While these efforts can reduce communication size, challenges still exist: the threshold of gradients to be discarded requires complex considerations. For example, keeping the communication size small (drop as many gradients as possible) can reduce each synchronization round's completion time, especially in a poorly performing network. However, the downside is that the additional computational overhead may be introduced, which may even outweigh the time optimization from reduced communications.

Among the above methods, Random-K packet loss and Top-K packet loss are the two most commonly used, and they have their own advantages and disadvantages. We use the model ResNet18 with the CIFAR10 dataset~\cite{krizhevsky2009learning} to Explore this issue. We guarantee k\% of the gradients to be synchronized and compare the differences in top-1 accuracy and throughput between randomly discarding 1-k\% of the gradients (Random-k) and keeping top k\% of the gradients (Top-k) in 8 worker nodes and one PS. We used CUDA's built-in topk function~\cite{cuda} to ensure the Top-k algorithm is efficient enough. The k value ranges from 5 to 40. Results (Figure~\ref{fig:randk-and-topk}) show that the Random-k can achieve relatively high accuracy and has higher training throughput due to its simplicity. When $k\leq 70$, the top-1 accuracy difference between Top-k and Random-k is only about 0.3\%, but there is about 25\% improvement in throughput. This provides guidance for the design of loss-tolerant transport protocols.


Other studies have proposed similar results~\cite{xu2021grace}. The LTP behaves as an approximate threshold-controlled Random-k, which means that the final result of DML training will not have an enormous impact due to \textit{limited} and \textit{random} data discarding.

\section{The Design of LTP}\label{sec:design}

LTP is a loss-tolerant transmission protocol for DML training under various network environments. We follow several fundamentals to design the LTP: 

\begin{enumerate}
    \item Ensure that the introduction of LTP has minimal impact on existing DML training synchronization.
    \item Mitigate the harm of long-tail latency caused by incast traffic patterns and Non-congestion packet loss.
    \item Reduce the impact of packet loss on the transmission window and avoid false link congestion signals caused by non-congestion reasons.
\end{enumerate}

This section will focus on these fundamentals and introduce the design concept and reasons for LTP.

\subsection{LTP Overview}


\begin{figure}[tbp]
\centerline{\includegraphics[width=9cm]{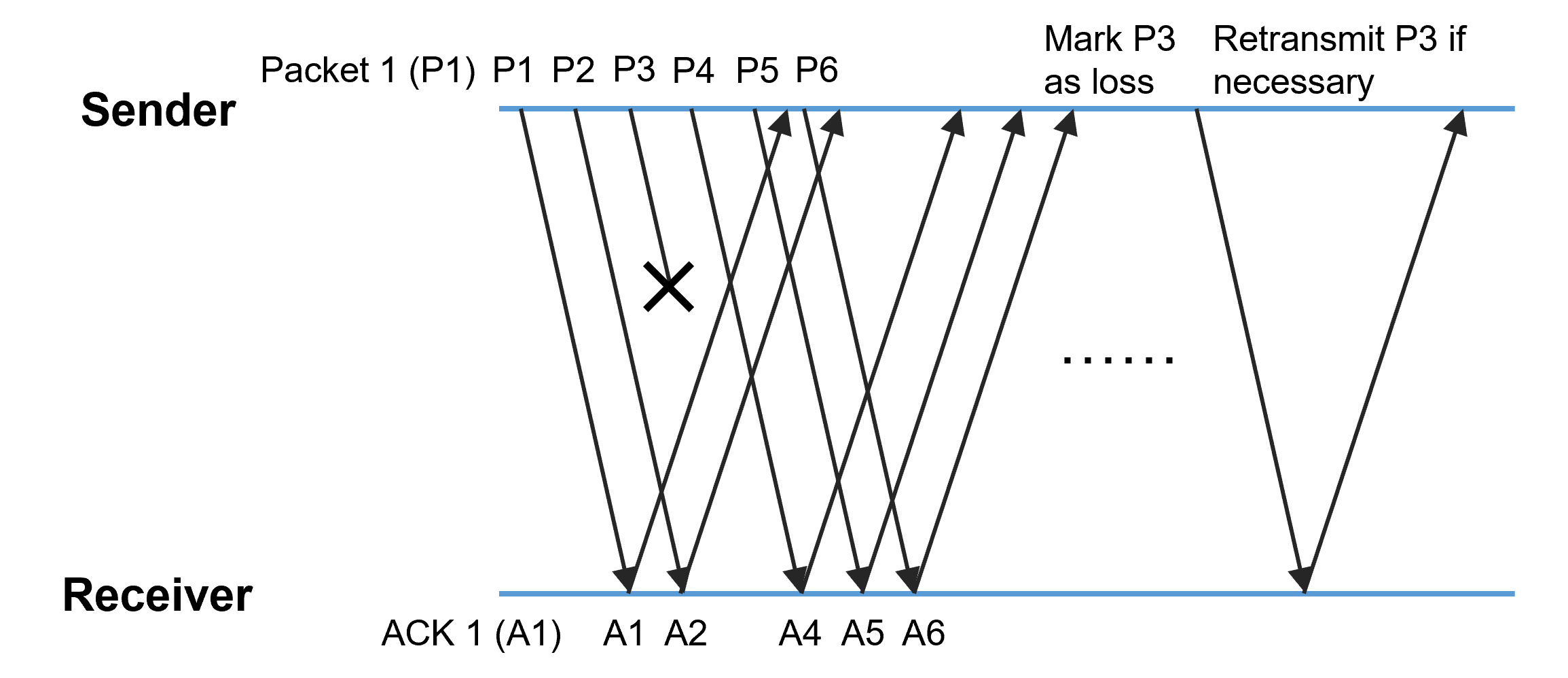}}
\caption{The loss-tolerant transmission in LTP.}
\label{figure:loss-tolerant_transmission}
\end{figure}

Different from TCP, the LTP utilizes two core solutions named \textbf{out-of-order packet transmission} and \textbf{per-packet ACK} to ensure loss-tolerant transmission. For TCP, the byte stream to be transmitted is split into multiple data segments, each of which is regarded as a separate piece of the byte streams to be delivered in order. However, order-preserving data transmission in DML is not required, so LTP utilizes out-of-order transmission and out-of-order ACK to improve protocol performance and support loss-tolerant transmission.

For example, in Figure~\ref{figure:loss-tolerant_transmission}, Packet 3 (P3) is marked as lost after three out-of-order ACKs. However, LTP does not retransmit P3 in real-time but waits for the completion of all packets sent before considering whether to retransmit. Whether the sender retransmits data at the end depends on the thresholds maintained by the Early Close mechanism(refer to \S~\ref{subsection:early_close}).

Here come several challenges to maintaining loss-tolerant transmission and keeping DML working properly. One of the most important challenges is to decide how much data should be transmitted. LTP proposes the Early Close mechanism based on real-time network quality that dynamically adjusts transmission thresholds, which can be referred to \S~\ref{subsection:early_close}. Another challenge is how to handle the data error caused by loss-tolerant transmission. To address this challenge, bubble-filling is proposed for data recovery on the receiver (\S~\ref{subsection:bubble_filling}). Other design details are also discussed in this section, including congestion control \textit{et al.}

\subsection{Cutting long-tail latency through Early Close} \label{subsection:early_close}

\begin{figure}[tbp]
\centerline{\includegraphics[width=9cm]{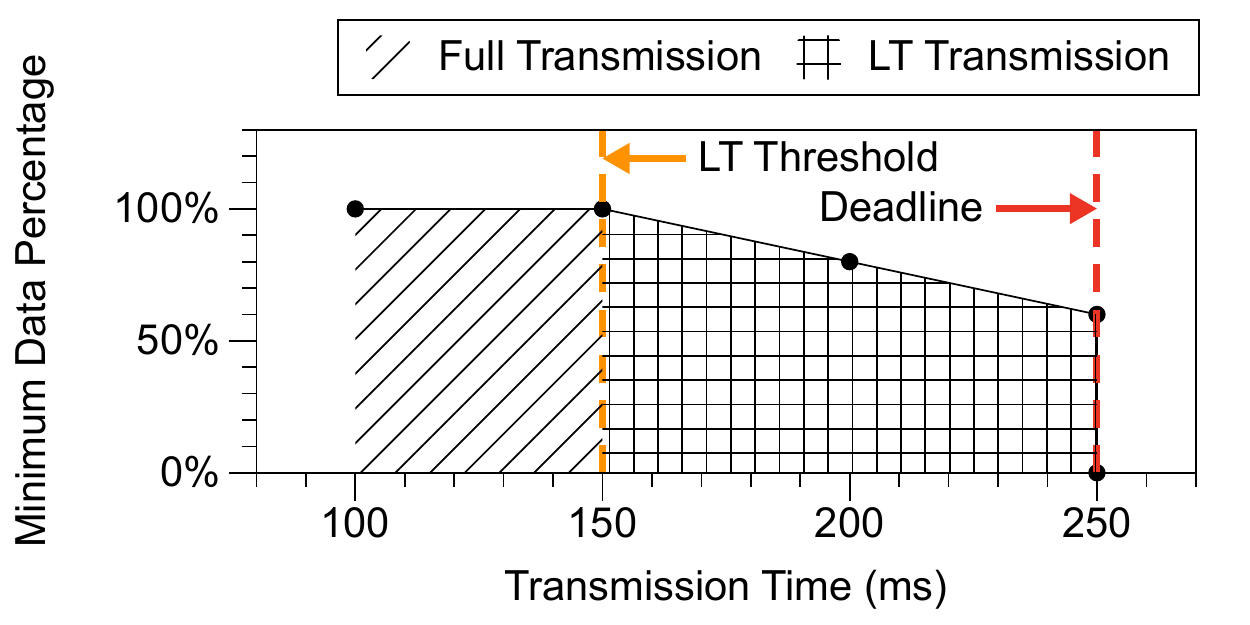}}
\caption{The double thresholds of Early Close mechanism.}
\vspace{-0.2in}
\label{fig:early-end}
\end{figure}

To avoid long-tail latency that slows down the global training time in the incast scenarios, we introduce the ``Early Close" mechanism. Early Close allows the receiver to close the flow once the transmission has received a certain percentage of data and all critical packets which contain metadata are confirmed to be delivered (\S~\ref{sec:packet-priorities}). The Early Close mechanism has two indicators: transmission time and received data percentage. 

Inspired by the Explicit Congestion Notification (ECN)'s double threshold design, we choose two thresholds on transmission time as the upper and lower bounds for one batch synchronization transmission. The smaller threshold is called the Loss-tolerant threshold (LT threshold), and the larger one is the deadline (Figure~\ref{fig:early-end}). Before the LT threshold, the receiver will wait until all the packets arrive. Between the two thresholds, the receiver ends the transmission early based on the received data percentage. For instance, in Figure~\ref{fig:early-end}, when the transmission has proceeded for 200ms, the transmission will be ended if the receiver (PS) has received larger than 80\% of the data from all senders (workers). After the deadline, the receiver stops receiving data immediately no matter how much data is received. The receiver will actively broadcast a ``stop'' message to the senders for the notification of Early Close.

\subsubsection{Update the Loss-tolerant Threshold}


In the Early Close mechanism, the configuration of thresholds is an important part. An appropriate LT threshold can significantly reduce the lag flow phenomenon caused by incast scenarios while ensuring little data is discarded. In practice, we find that configuring the LT threshold value to the excepted completion time (ECT) of the gradients to be transmitted has better performance. The ECT can be calculated as $ECT = RTprop + ModelSize / BtlBw$.

However, packet loss may result in errors in the bottleneck bandwidth (BtlBw) and round-trip propagation time (RTprop) measured by the LTP's congestion control (\S~\ref{subsection:cc}), leading to the ECT being smaller than the time required for the model transmission. Although the threshold can be shared between different epochs (the flow size remains the same for the fixed number of gradients to be updated), adjustment is still needed because other competing flows in the network are dynamic. Combining these cases, we set the initial value of LT Threshold to  $LTThreshold_{init} = 1.5 * RTprop + ModelSize / BtlBw$ at the first batch of each epoch.

LT threshold is updated by the shortest 100\% gradient transmission time during every epoch. The LT Threshold is independent between each point-to-point link and adjusted by the optimal synchronization time of the current link. The deadline is applied on all receiving links of one receiver at the same time and takes the value as $Max(LT Threshold) + C$. The constant $C$ is user-defined. We set it to 30ms in DCN and 100ms in WAN. These values are empirical values, which have better performance in the evaluations.

\subsubsection{Different LT Thresholds when Gathering and Broadcasting}

LTP avoids the long-tail latency caused by incast traffic by the Early Close mechanisms. In practice, there are two processes for DML training in the PS architecture, gathering (the worker node sends its training gradients to the PSes) and broadcasting (the PSes aggregate the gradients and then send them back to the workers). These two processes have different tolerance for data loss. What is obvious is that the gathering process allows data loss, while the broadcasting process does not need because of 1) the characteristics of DML itself and 2) the incast traffic model caused by the PS architecture, respectively.

In terms of the characteristics of DML training, the acceleration brought by DML training is to split the dataset by multiple computing nodes and then calculate in parallel at the same time. Therefore, we believe that ML models among different machines should be consistent to avoid global model confusion. For instance, the DML task using Asynchronous Parallel synchronous models (ASP)~\cite{recht2011hogwild} suffers from the problem of low final training accuracy or even failure to converge because it cannot guarantee the synchronization of training models among worker nodes. As a result, the loss-tolerant transmission works only during the gathering but ensures all data are transmitted in broadcasting.

From the perspective of traffic patterns, the traffic in the broadcasting stage is not the incast pattern with severe resource competition but a one-to-many pattern. This traffic pattern will not cause long-tail latency and thus does not require the Early Close mechanism.

\subsection{Use Bubble-filling to Avoid Data Error} \label{subsection:bubble_filling}

\begin{figure}[tbp]
    \centering
    \subfigure[Problems brought by breaking floating point numbers.]{\includegraphics[width=\columnwidth]{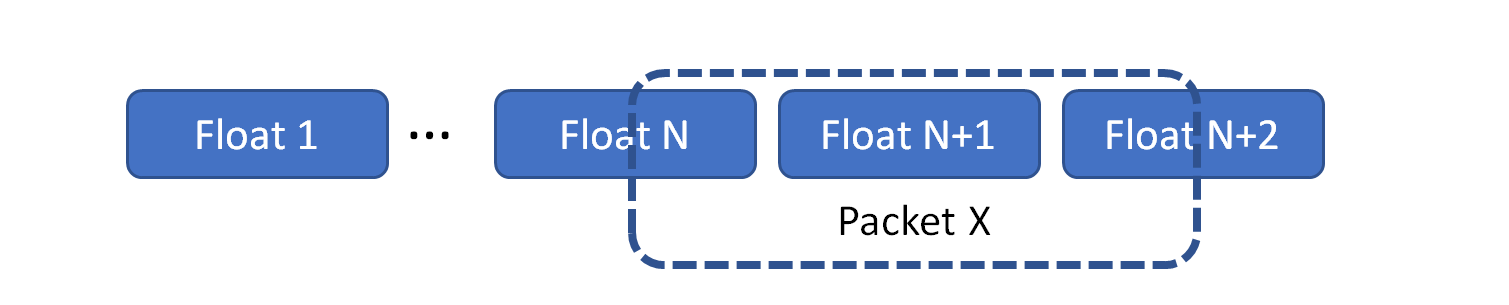}\label{subfigure:breaking_fp}}
    \subfigure[The padding bubbles can avoid the break of floating point numbers.]{\includegraphics[width=\columnwidth]{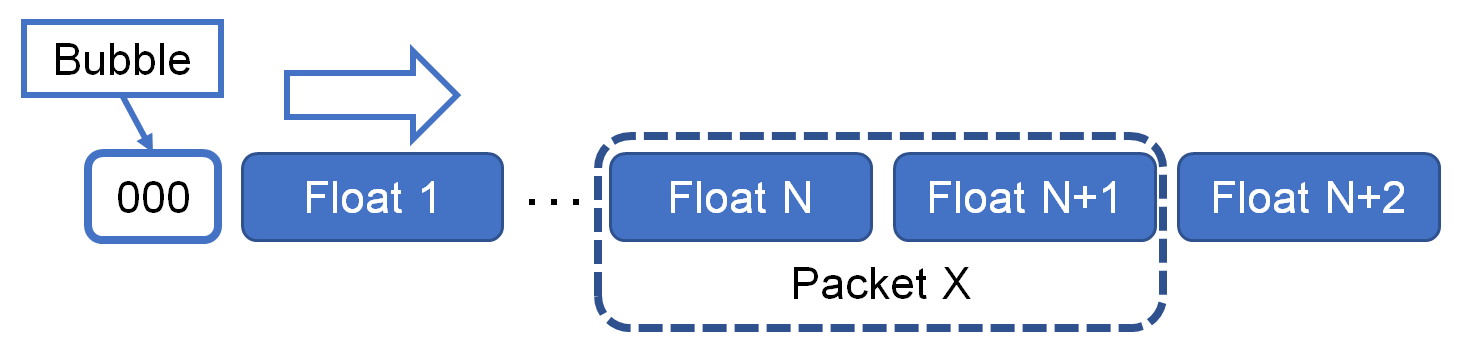}\label{subfigure:padding_bubble}}
    \caption{Bubble-filling can be used to correct severe data errors caused by breaking floating point numbers.}
    \vspace{-0.1in}
    \label{figure:bubble_filling_all}
\end{figure}


Since LTP is designed to support loss-tolerant transmission, packets being discarded are acceptable without being retransmitted. These lost packets can lead to serious data errors, for which we propose the bubble-filling mechanism to fix the problem caused by loss-tolerant transmission. In detail, bubbles are a variable-length string of all zeros, and two bubbles are used for correctness, named ``packet bubble'' and ``padding bubble''. 

For those packets that the LTP receiver believes should not be (re)transmitted, the receiver uses packet bubbles to fill them. In this case, these bubbles have the same length as the data packet. Since the Maximum Transmission Unit (MTU) will not change during transmission, the receiver can deduce the length of the packet bubble from the context packets.

However, in some cases, such an operation can cause the receiver to receive the wrong segmentation. For instance, when a single floating number is split into two segments, and only one of the segments of this floating number is filled by 0, this can result in an error in the value on the receiver, as shown in Figure~\ref{subfigure:breaking_fp}. The good news is that most of the gradients to be transmitted in DML training are aligned with the same type, such as float32 (occupying the same memory spaces). Therefore, LTP uses the padding bubble (Figure~\ref{subfigure:padding_bubble}) to make sure that there is no wrong segmentation. 

\subsection{Congestion Control} \label{subsection:cc}

In both DCNs and WANs, the DML training task competes with other flows for link bandwidth, switch queues, and NIC buffers. Discussing the trade-off between fairness and efficiency among different traffics is required. Different congestion control algorithms are usually proposed to deal with this problem. Traditional TCP congestion control adjusts the sending rate through the additive-increase/multiplicative-decrease (AIMD) of the congestion window size, ensuring that data arrives intact and in order by using 3 duplicated ACKs as packet loss signals. However, non-congestion packet loss in WANs may trigger false congestion signals and prevent full link utilization. In contrast, BBR is a novel congestion control algorithm that performs well in networks with non-congestion packet loss, by using BDP as the upper limit of the number of packets in flight and pacing to avoid buffer overflow.

LTP's congestion control algorithm is BDP-based and takes effect at the sender, in which the recognition of packet loss is not used as a signal to adjust the cwnd. LTP estimates BDP by periodically probing the RTprop and BtlBw, respectively, and uses BDP as the maximum count of packets in flight. We can not use BBR directly because it is built on top of TCP and is harder to modify. Therefore, the pacing can not be performed precisely because LTP runs in the user space (UDP). We use an approximate pacing scheme where we execute a wait function based on the pacing rate calculated by BBR when the count of packets to be sent at the same time is greater than 20 (10G link, MTU 1500 Bytes, ~30KB) (this is not a common situation when the sending window is stable).

\subsection{Packet Priorities }\label{sec:packet-priorities}

Many gradient compression approaches in DML discuss the importance of how to rank different data. In LTP, data importance is equally worth discussing. LTP transmits different packets with different reliability by dividing them into two priorities: 1) critical packets and 2) normal packets. LTP determines that all critical packets are 100\% received by the receiver during transmission, while normal packets can be partially dropped. LTP allows users to customize the selection of critical packets while marking only a minimal number of messages as critical.

LTP marks only those indispensable bytes of the matrix (e.g., several bytes on the first and last part of the matrix bitstream). LTP does not traverse the specifics of what the data in the matrix represent, such as the absolute value of each data, because 1) traversing the whole matrix is time-consuming and 2) ranking the importance based on the absolute value of the data is not ideal (refer to \S~\ref{sec:selection-of-dropped-gradients}).

\section{Implementation of LTP}\label{sec:implementation}

\begin{figure}[tbp]
\centerline{\includegraphics[width=0.9\columnwidth]{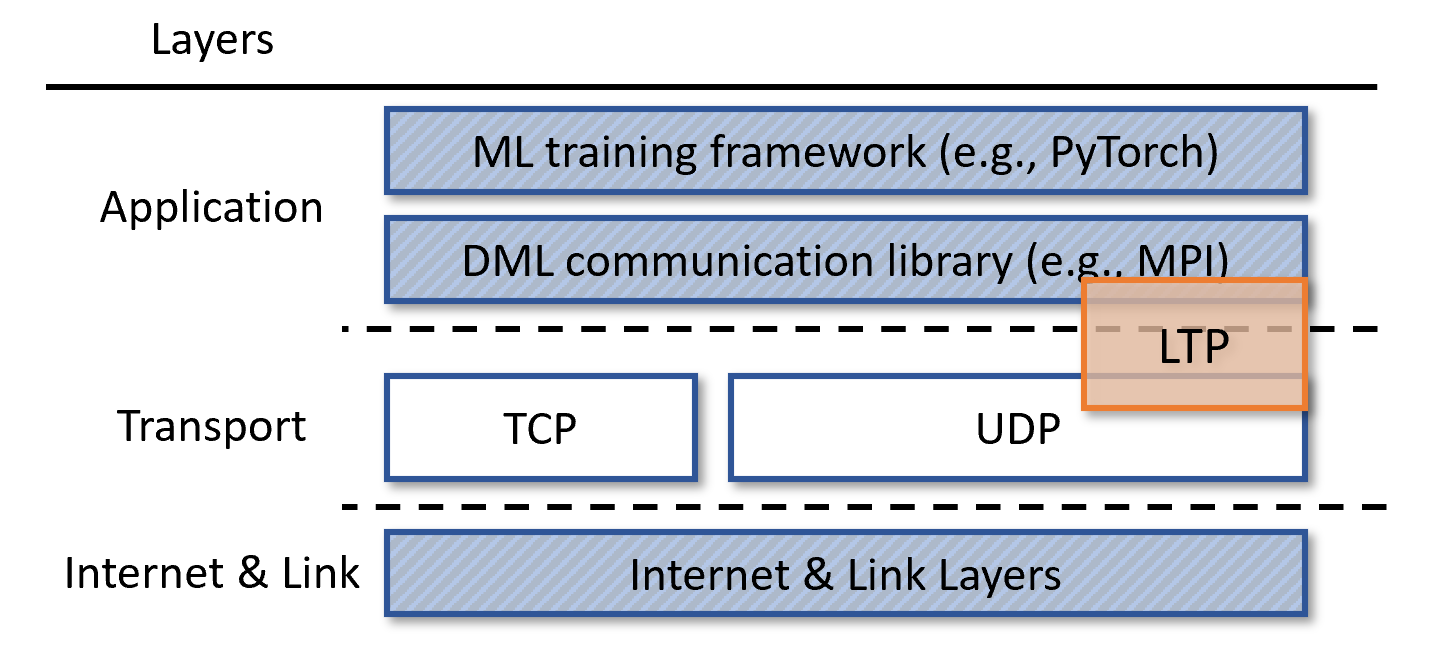}}
\caption{LTP is designed based on UDP. Therefore, it is located primarily at the transport layer and includes some application layer functionality.}
\label{fig:ltp-layer}
\end{figure}

We deploy the LTP protocol on Linux (Ubuntu 18.04.2) using C++ and integrate it into the existing widely used DL framework PyTorch. LTP is deployed based on UDP with modifications on the DML communication library (Figure~\ref{fig:ltp-layer}). This section describes the design details of the implementation of LTP, including the data structure of the LTP packet and how to perform queue management.

\subsection{Data Structure}

\begin{figure}[tbp]
\centerline{\includegraphics[width=0.9\columnwidth]{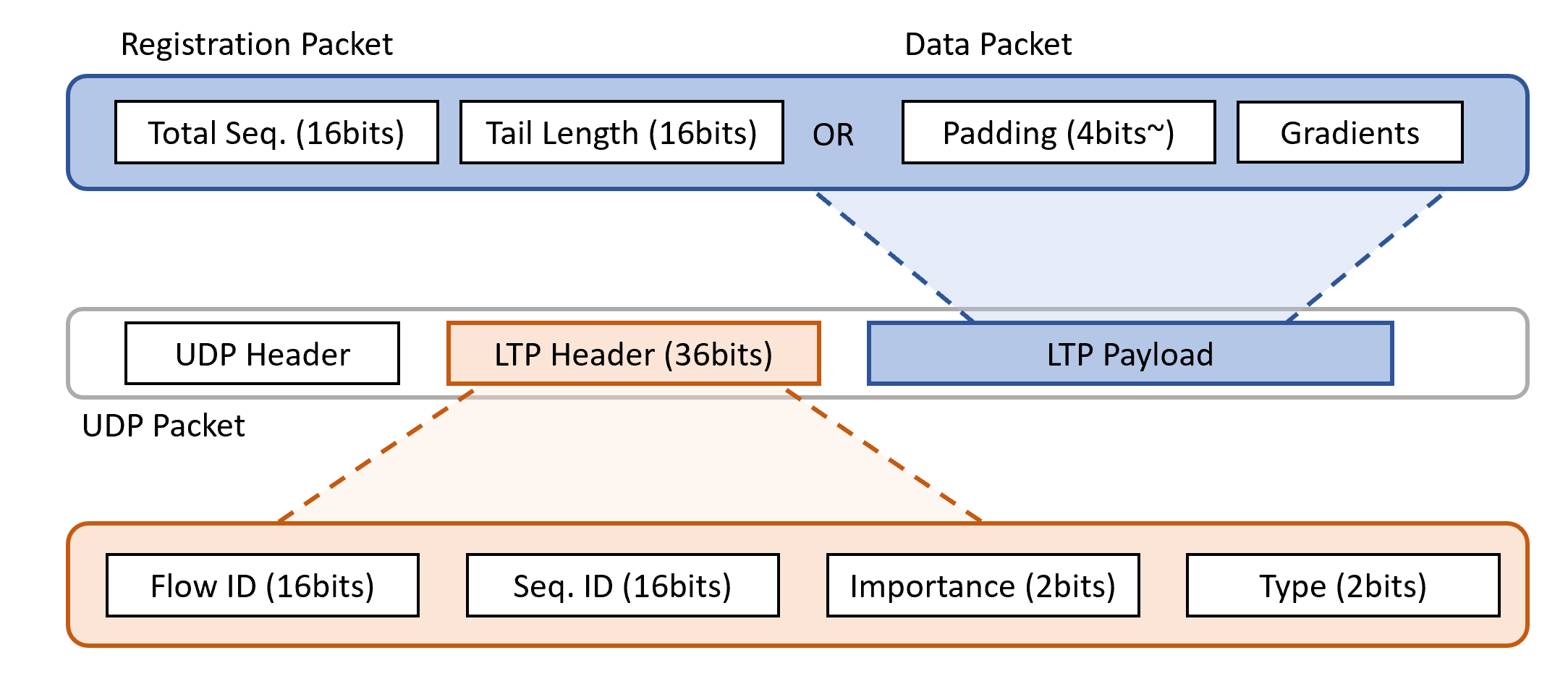}}
\caption{LTP Packet Structure.}
\label{fig:packet-structure}
\end{figure}

LTP redesigns the data structure of the packet (Figure~\ref{fig:packet-structure}). LTP runs over UDP, which is similar to most of the custom transmission protocols (\textit{e.g.}, QUIC~\cite{iyengar2020quic}, RoCEv2~\cite{infiniband2014rocev2}).

To optimize the packet transmission efficiency, LTP only adds a header of additional 68 bits (about 9B). LTP header contains the fields of the flow ID, the sequence ID, the importance of the packet, the packet type, the RTprop, and the BtlBw. Each round of transmission is regarded as a separate flow. The receiver distinguishes between different flows by recording the sender's address and the flow ID. The sequence ID represents the order of the piece of the Jigsaw (data segment). The importance field is used to mark the significance of the packet; in the initial design, the importance field contains only two categories, critical (11) and not critical (00). The type field is used to mark the property of the packet, including registration packet (00), data packet (01), ACK (10), and end (11). LTP sends the congestion control information to the receiver, containing the RTprop and the BtlBw.

The payload field has different roles in different types of messages. The payload field holds the total number of data segments for the specific flow in registration packets. In data packets, it is used for keeping the segmented data. The payload field stays empty in ACK packets.

\subsection{Queue Implementation}\label{sec:queue_management}

\begin{figure}[tbp]
\centering
\subfigure[The CQ guarantees the arrival of all packets (similar to classical TCP).]{\includegraphics[width=\columnwidth]{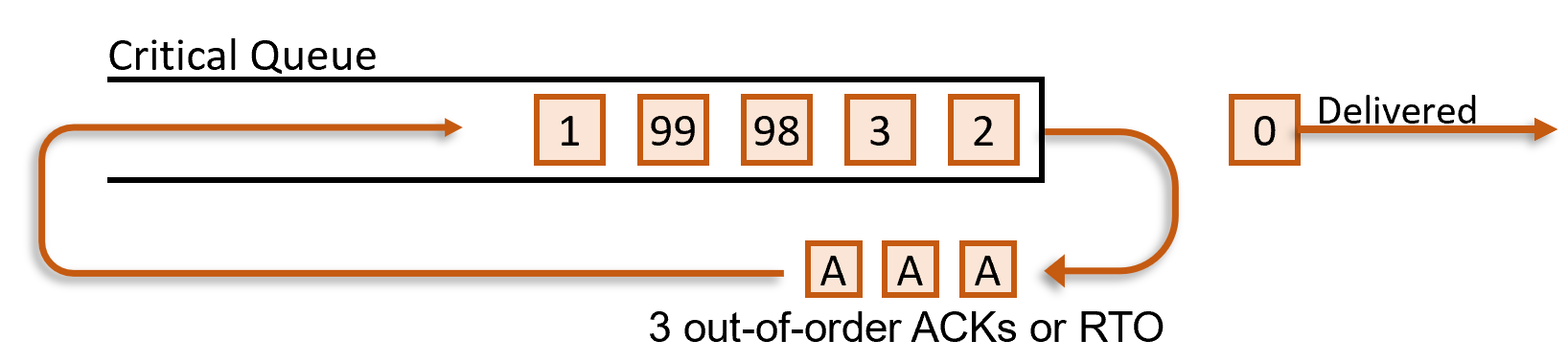}
\label{subfigure:cq}}
\subfigure[Packets in the NQ are transmitted only once. When a lost packet is detected, it will be retransmitted via the RQ.]{\includegraphics[width=\columnwidth]{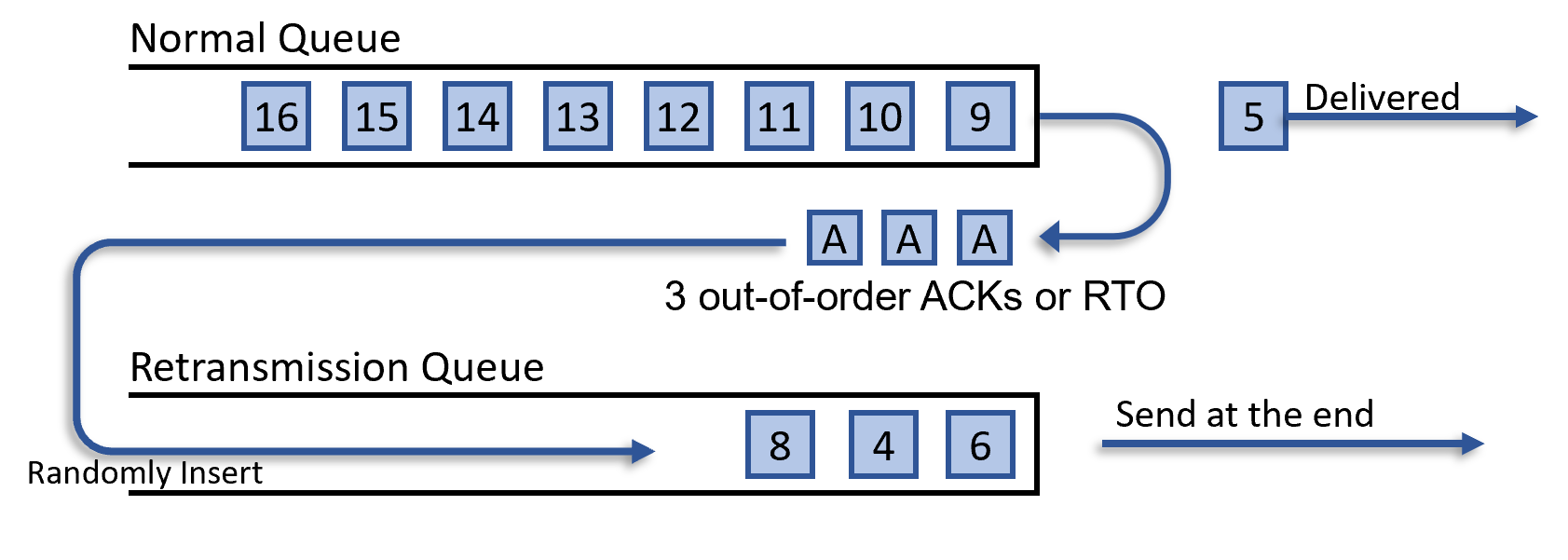}
\label{subfigure:nq}}
\caption{Detailed queue management in LTP.}
\vspace{-0.1in}
\label{fig:queue-implementation}
\end{figure}

To ensure different transmission orders of packets with different priorities, we design 3 queues. Two of them are First-in, First-out (FIFO) queues for the delivery of packets, namely the \textbf{Critical Queue (CQ)} (Figure~\ref{subfigure:cq}) and the \textbf{Normal Queue (NQ)}  (Figure~\ref{subfigure:nq}). The other queue is \textbf{Retransmission Queue (RQ)}, which is a Random-in, First-out queue. 

LTP splits the byte stream to be delivered into packets, then placed in each of the two queues according to the rules. The packets in NQ are sent after CQ, and the packets recognized as lost are reinserted into CQ (packets in CQ) or RQ (packets in NQ) to wait for retransmission. RQ, which is used to store those packets recognized as lost, will start sending packets when all the messages in CQ and NQ have been transmitted. 

LTP recognizes the packet loss by the three out-of-order ACKs. LTP maintains a queue of actual packet outgoing sequences on the sender, which is used as the basis for considering the ACK's arrival order.


\begin{figure*}[htbp] 
\centering  
\includegraphics[width=14
cm]{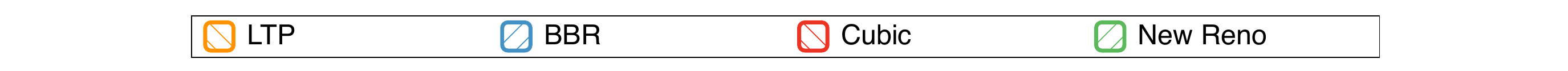}

\subfigure[ResNet50]{
\includegraphics[height=3cm]{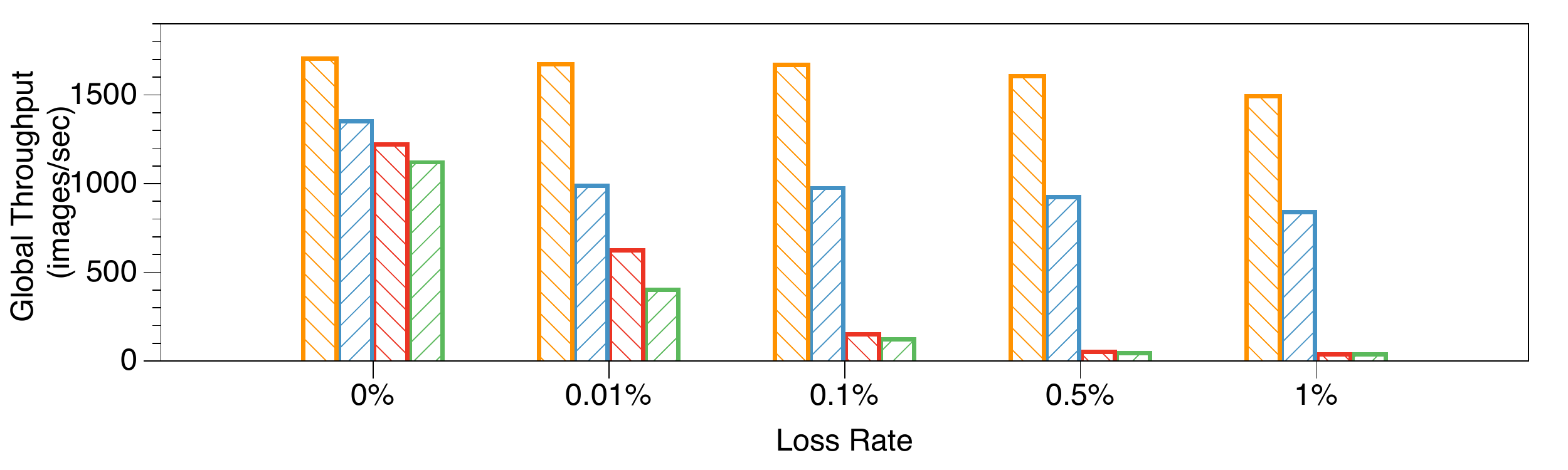}
}
\subfigure[VGG16]{
\includegraphics[height=3cm]{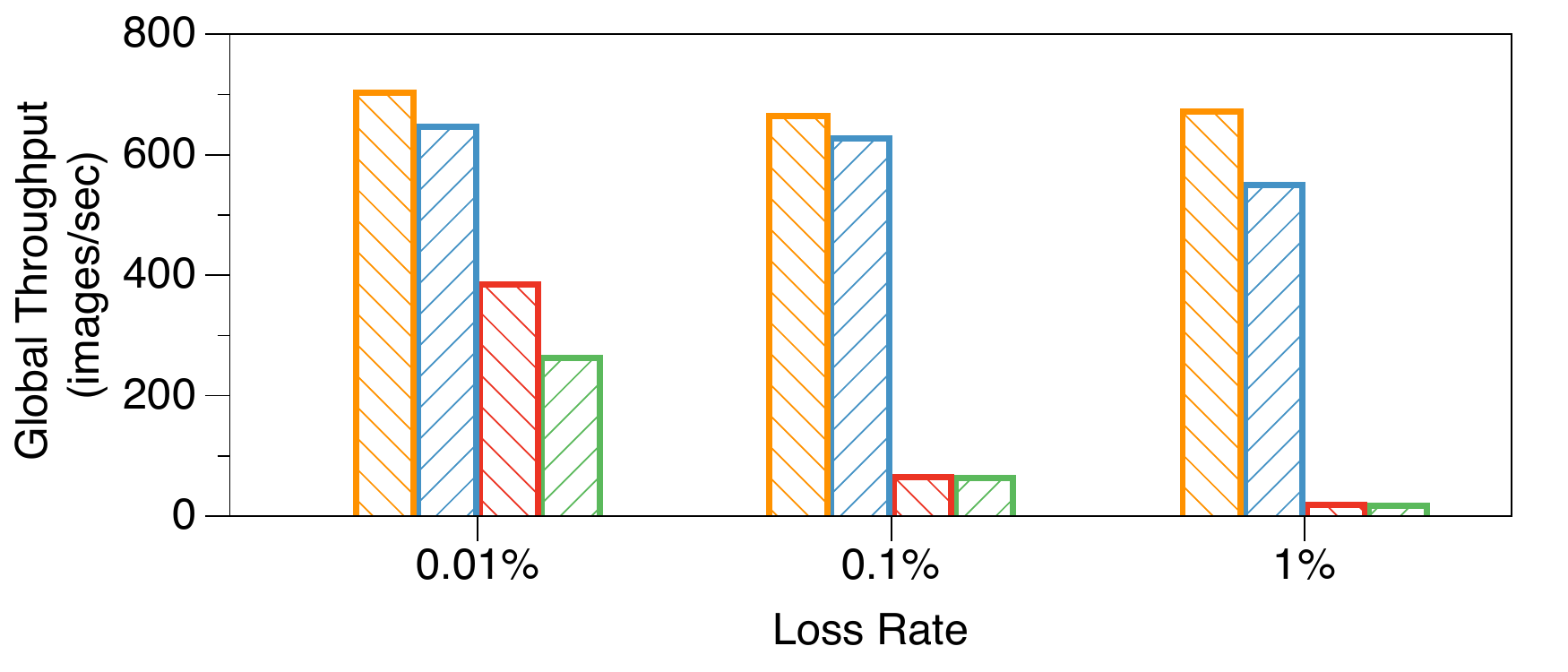}
}

\caption{Throughput of different congestion control algorithms under different non-congestion loss rates}
\label{fig:throughput}
\end{figure*}

\section{Evaluation}\label{sec:evaluation}

We use testbed experiments to evaluate LTP from these perspectives:

\begin{enumerate}
    \item How much of a throughput improvement does LTP provide to existing DML training tasks? (\S~\ref{sec:throughput_and_tta})
    \item How does LTP perform in a network with non-congestion packet loss? (\S~\ref{sec:bst})
    \item Will bubble-filling lead to low convergence accuracy at the end? (\S~\ref{sec:throughput_and_tta})
    \item How does LTP perform in terms of fairness when coexisting with other transmission protocols? (\S~\ref{sec:fairness})
\end{enumerate}

\begin{figure*}[htbp] 
\centering  
\includegraphics[width=14
cm]{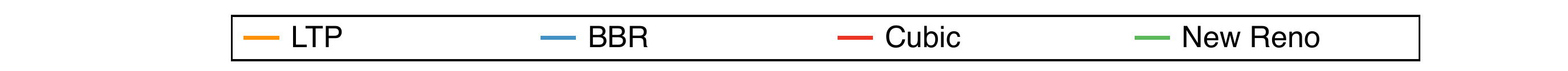}

\subfigure[ResNet50, loss rate: 0\%]{
\includegraphics[width=0.45\columnwidth]{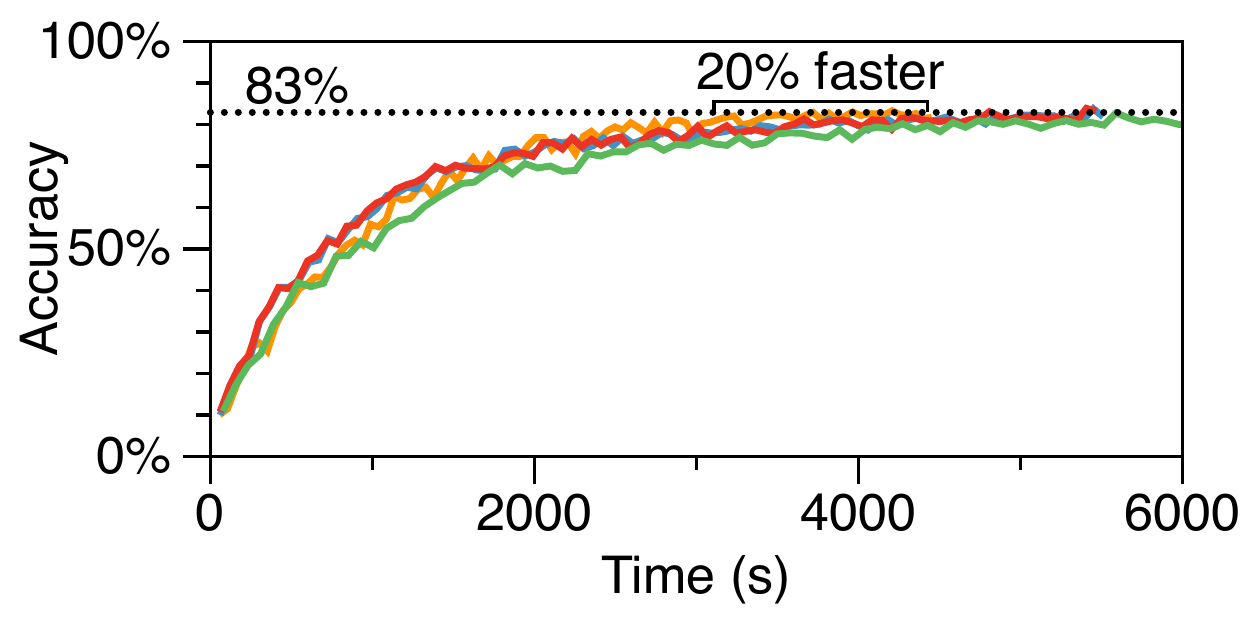}
}
\subfigure[ResNet50, loss rate: 0.01\%]{
\includegraphics[width=0.45\columnwidth]{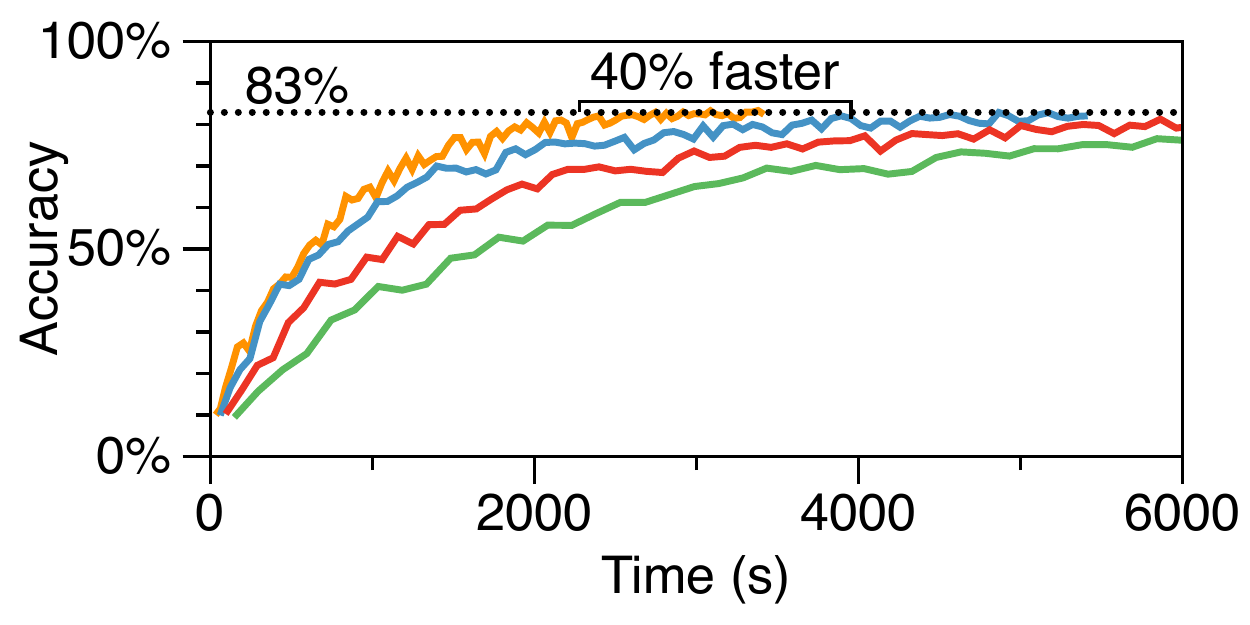}
}
\subfigure[ResNet50, loss rate: 0.1\%]{
\includegraphics[width=0.45\columnwidth]{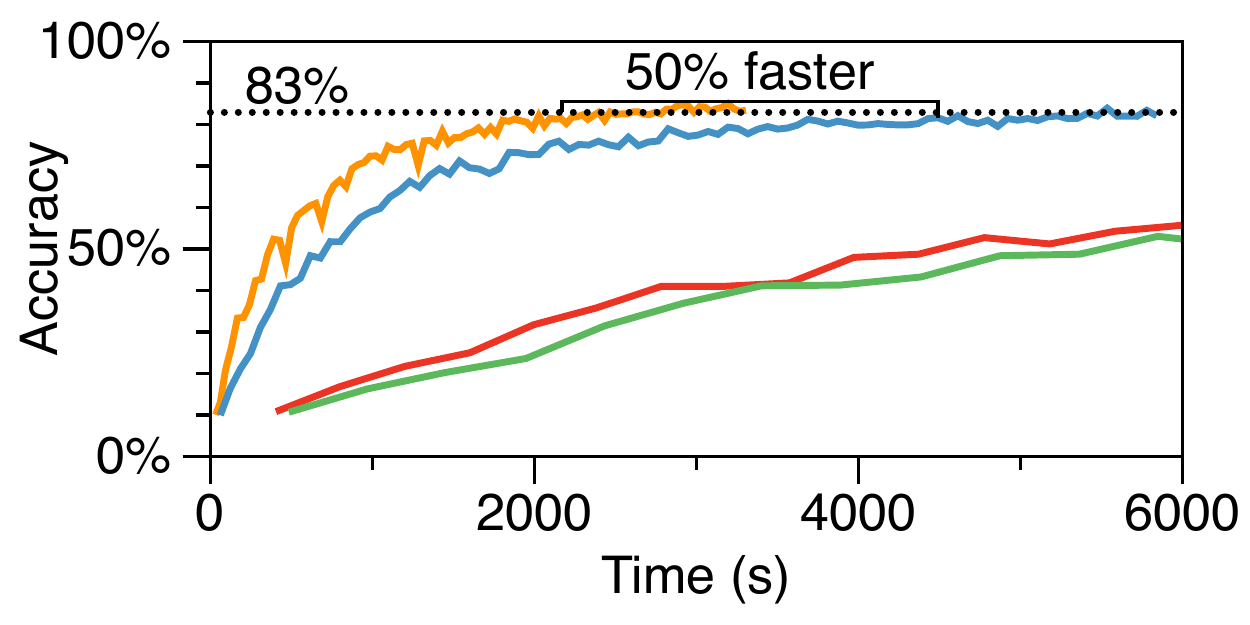}
}
\subfigure[ResNet50, loss rate: 0.5\%]{
\includegraphics[width=0.45\columnwidth]{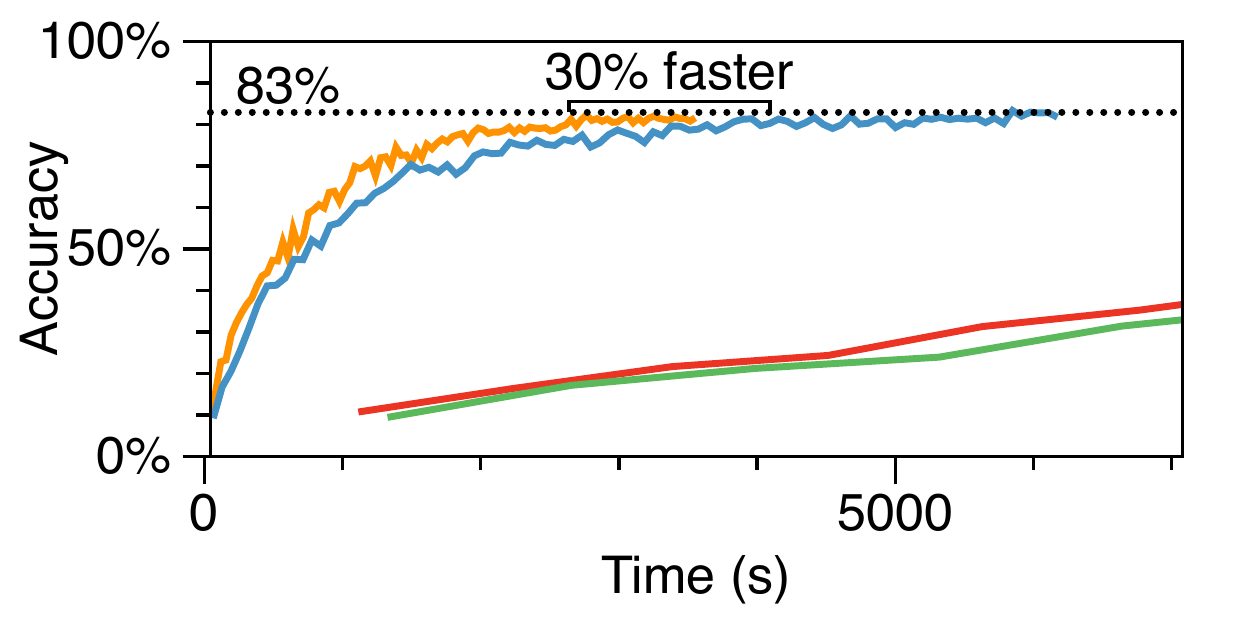}
}
\subfigure[ResNet50, loss rate: 1\%]{
\includegraphics[width=0.45\columnwidth]{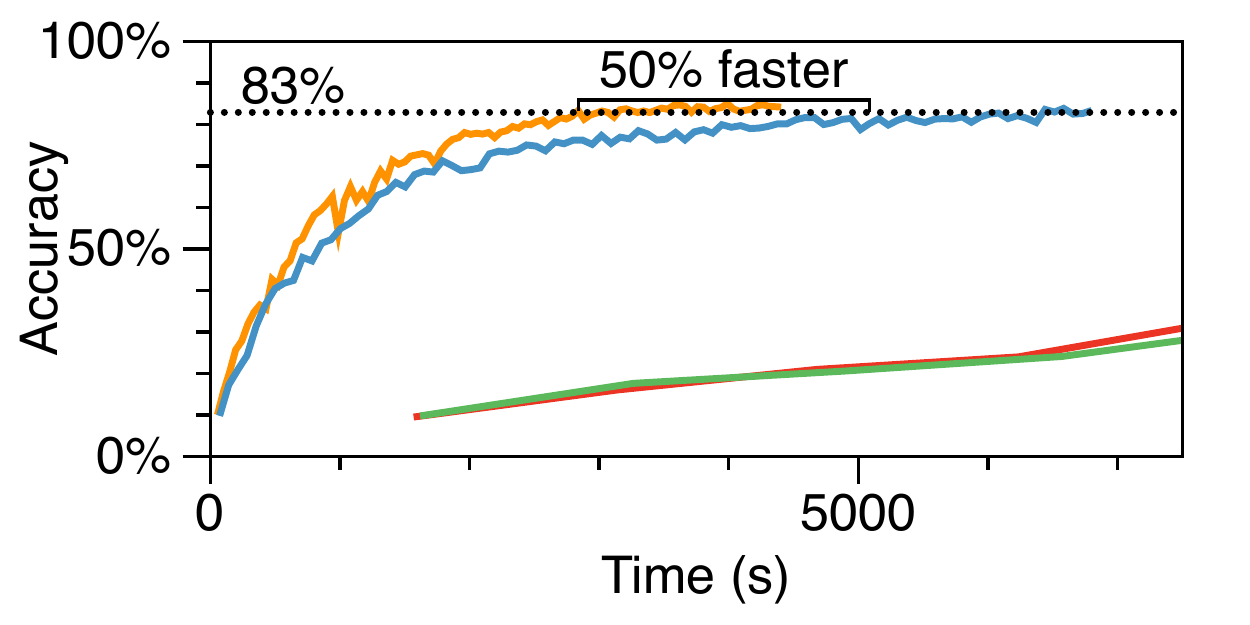}
}
\subfigure[VGG16, loss rate: 0.01\%]{
\includegraphics[width=0.45\columnwidth]{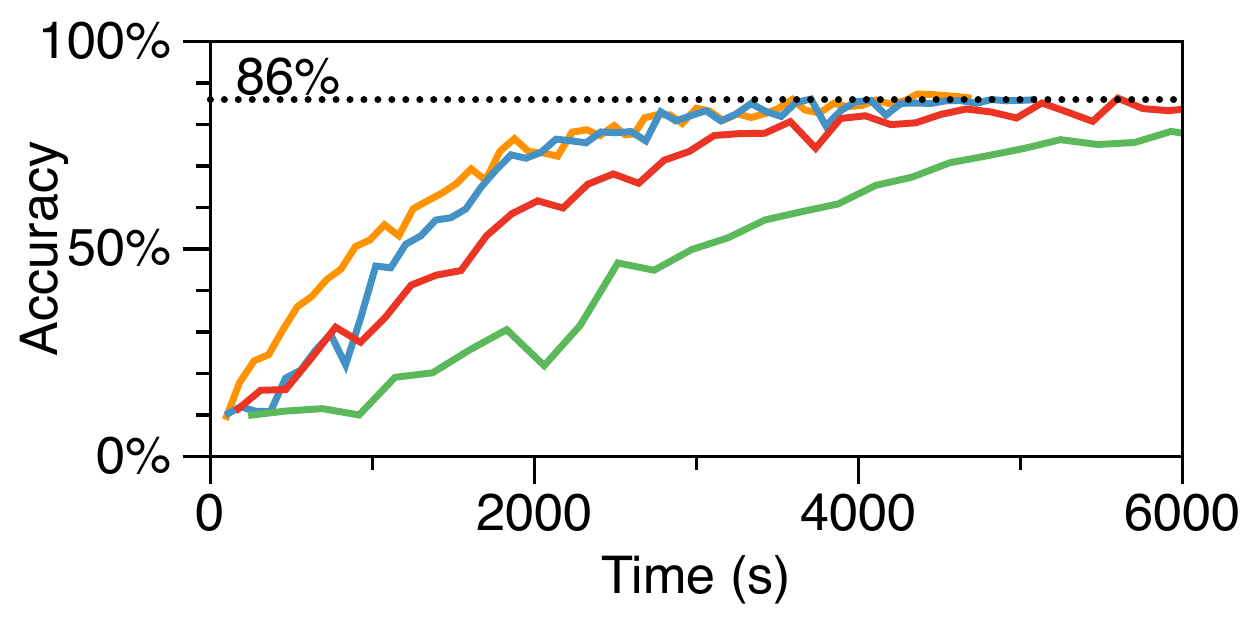}
}
\subfigure[VGG16, loss rate: 0.1\%]{
\includegraphics[width=0.45\columnwidth]{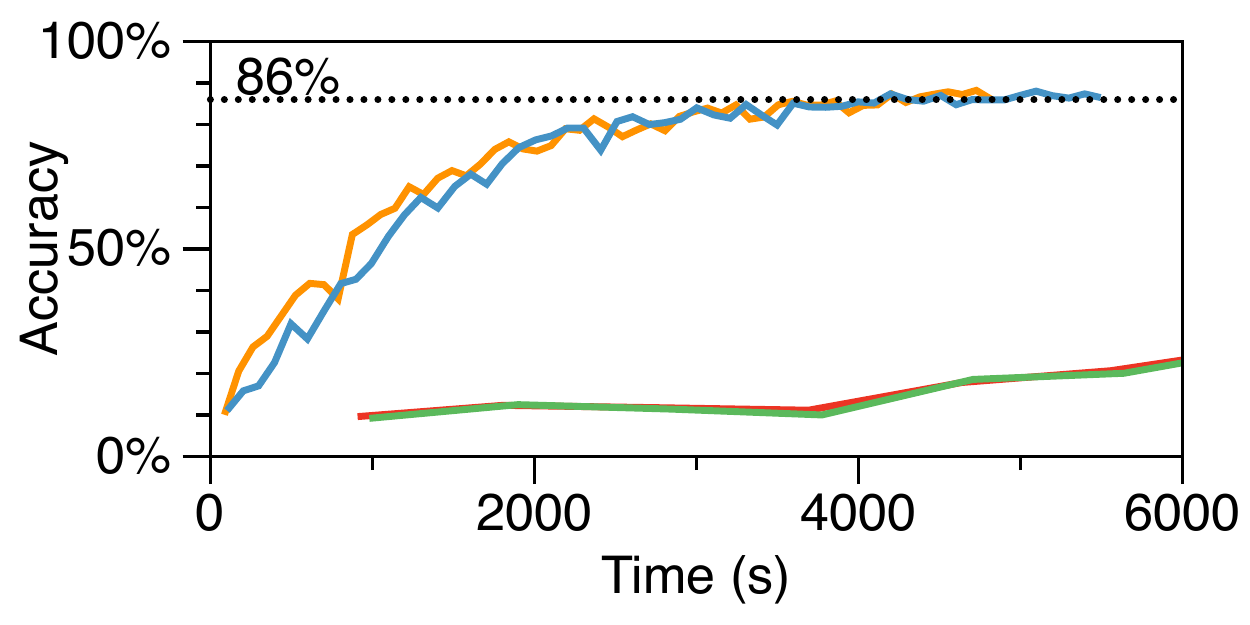}
}
\subfigure[VGG16, loss rate: 1\%]{
\includegraphics[width=0.45\columnwidth]{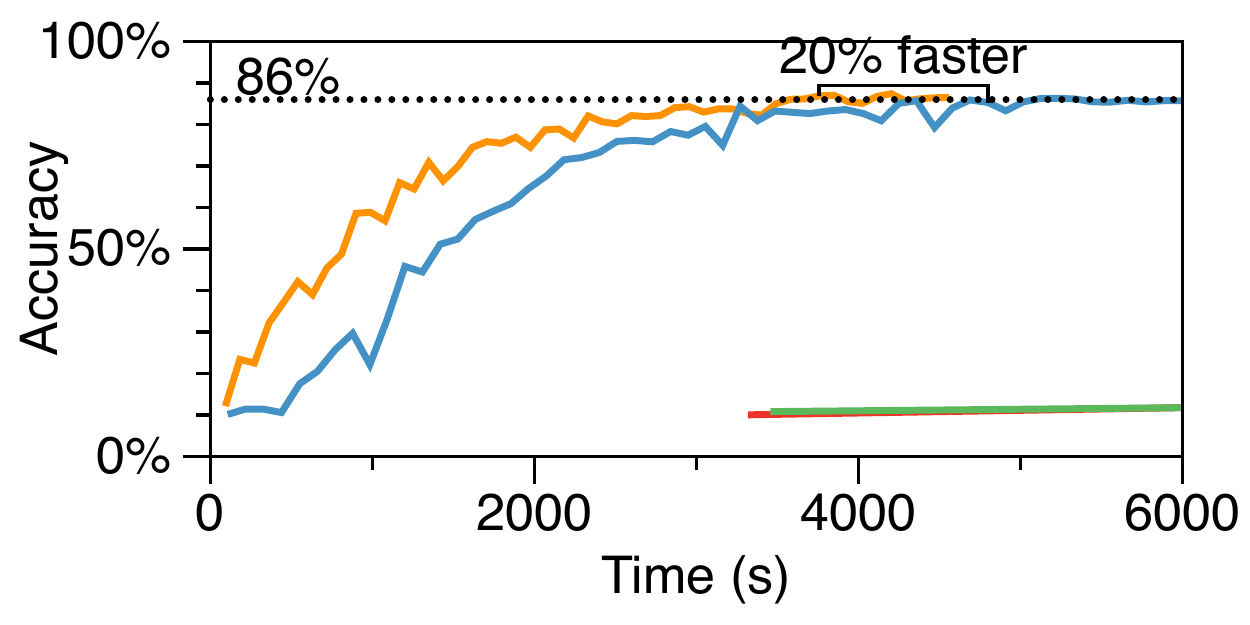}
}
\caption{TTA between different congestion control algorithms under different non-congestion loss rates.}
\label{fig:time_to_accuracy}
\end{figure*}

\begin{figure*}[htbp] 
\centering

\subfigure[ResNet50, LR: 0\%]{
\includegraphics[width=0.35\columnwidth]{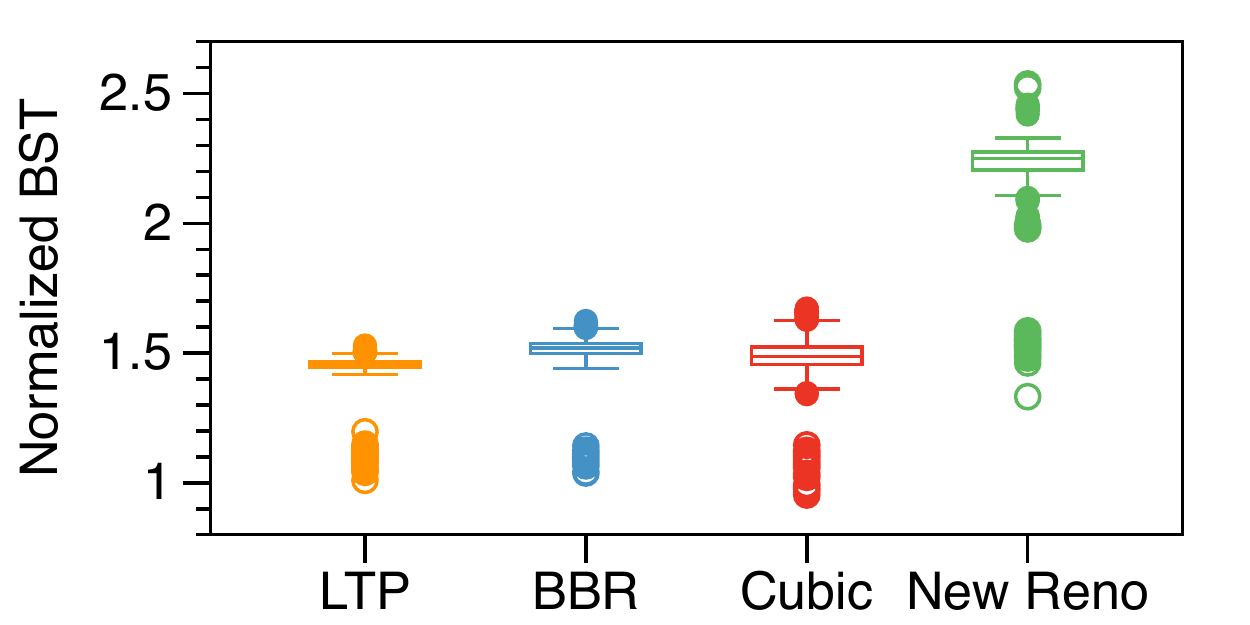}
}
\subfigure[ResNet50, LR: 0.01\%]{
\includegraphics[width=0.35\columnwidth]{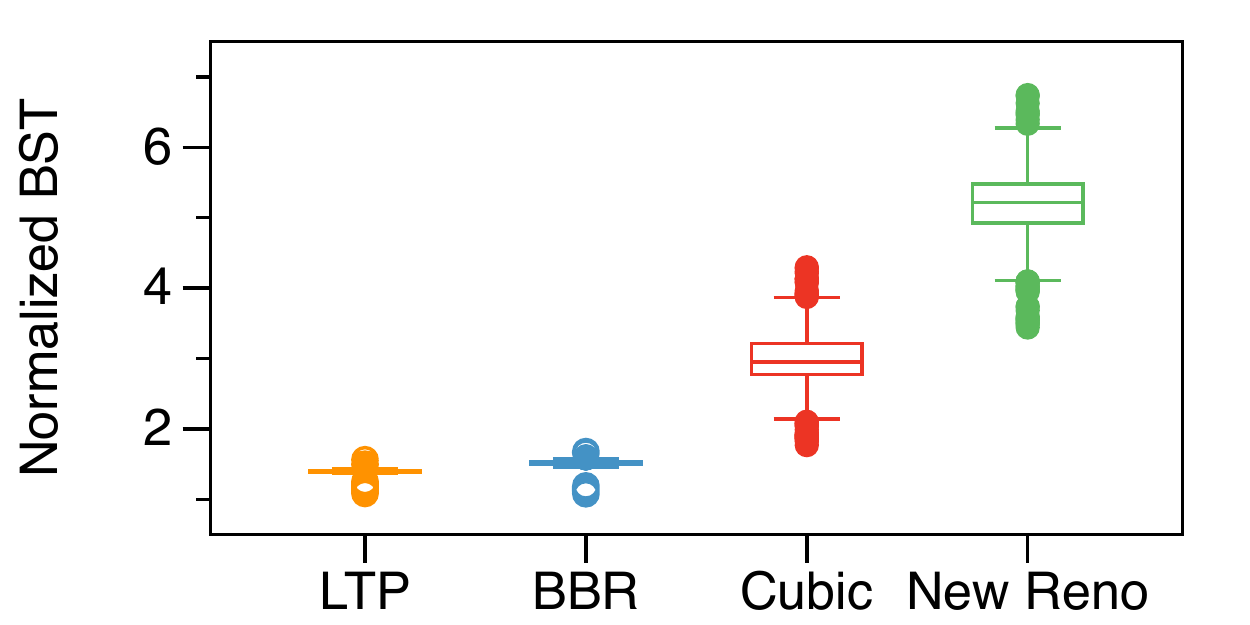}
}
\subfigure[ResNet50, LR: 0.1\%]{
\includegraphics[width=0.35\columnwidth]{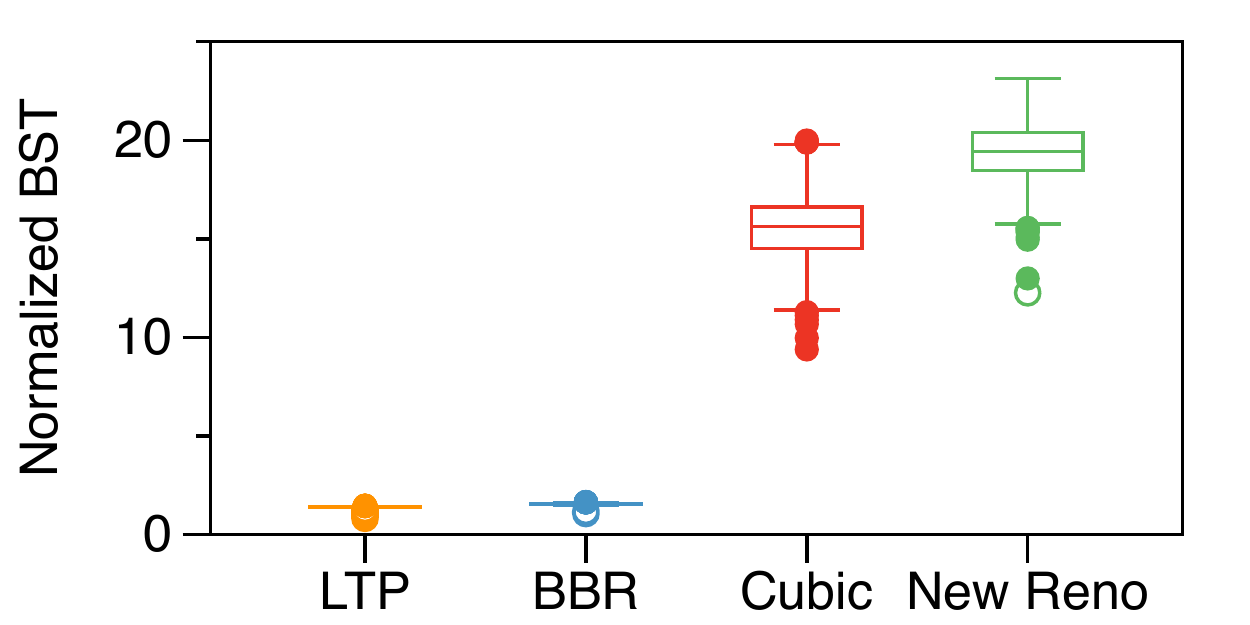}
}
\subfigure[ResNet50, LR: 0.5\%]{
\includegraphics[width=0.35\columnwidth]{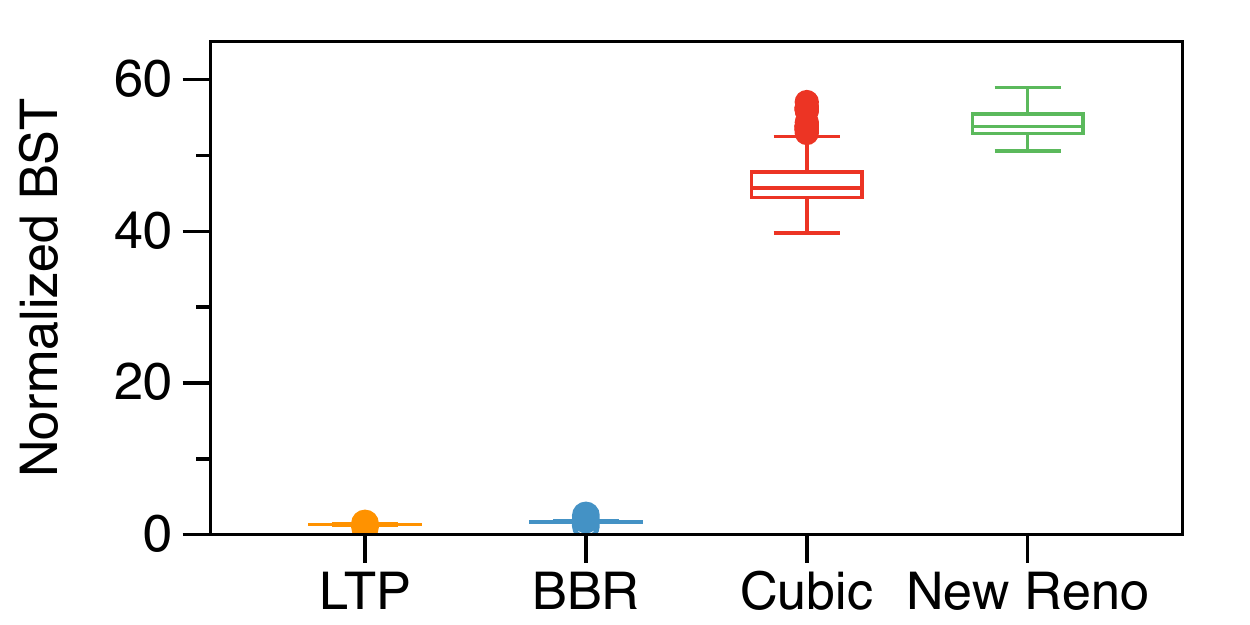}
}
\subfigure[ResNet50, LR: 1\%]{
\includegraphics[width=0.35\columnwidth]{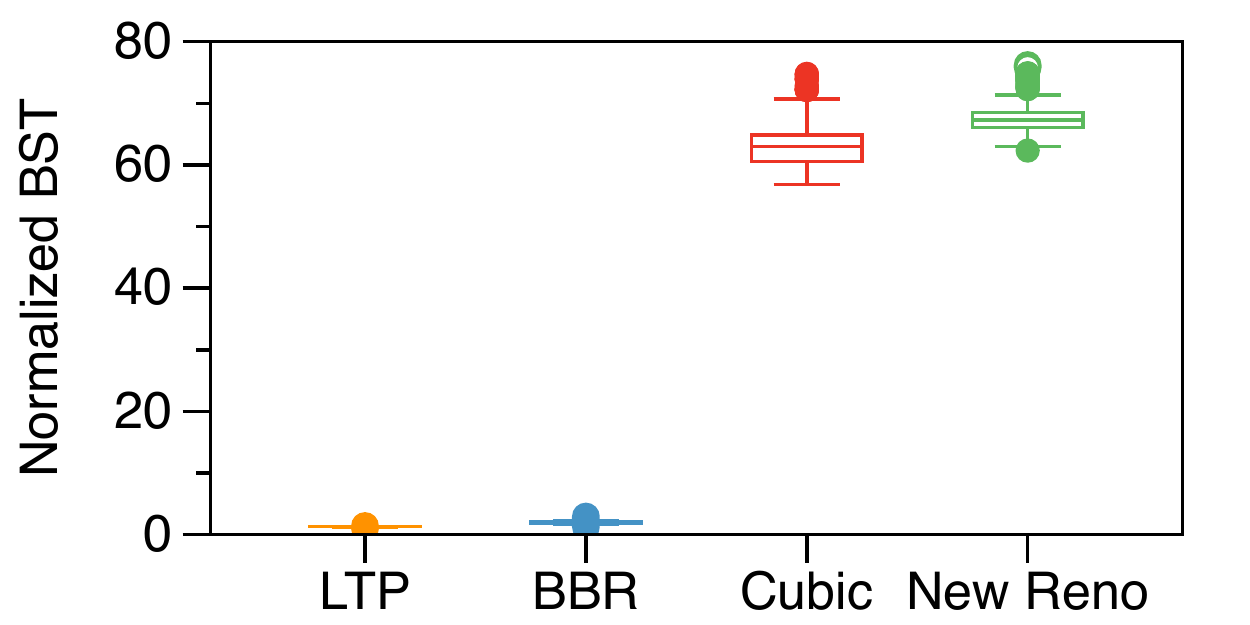}
}
\caption{BST (Normalized to LTP) on ResNet50 between different congestion control under different non-congestion loss rates.}
\vspace{-0.1in}
\label{fig:batch_time}
\end{figure*}

\subsection{Experiment Setup}

\subsubsection{Cluster Configurations}

We evaluate the LTP on a cluster with 9 machines. One of the machines is used as the PS, and the other 8 are worker nodes. All the nodes are located in one rack and connected to a HUAWEI CE6881-48S6CQ Layer-3 ToR switch. Each machine has 2 Intel(R) Xeon(R) Silver 4116 CPU @ 2.10GHz (24 cores, 48 threads), 256GB RAM, and one Intel(R) Ethernet Network Adapter X722. Each machine is also equipped with an NVIDIA Tesla T4 16GB GPU with NVIDIA driver version 460.91.03, and CUDA 11.2~\cite{cuda}. The OS of each machine is Ubuntu 18.04.2 with kernel version 4.15.0. 

\subsubsection{Baselines}

Since we only modify the transmission protocol and make no changes to the other logic of DML training, to control the variables, we design our own PS-based DML framework for all evaluations we perform. We use the TCP congestion control algorithm with default parameters from the Linux kernel as a baseline for comparison with LTP (LTP uses UDP and is not affected by the kernel congestion control algorithm). We evaluate LTP in comparison with existing widely used TCP congestion control algorithms, including BBR\cite{cardwell2016bbr}, Cubic\cite{ha2008cubic}, and New Reno\cite{floyd2004newreno}. The MTU is set as 1500, and we adjust the QoS policy on the ToR switch to ensure that it treats TCP and UDP equally.

\subsubsection{Workloads}

We run PyTorch on the cluster and the framework above to perform the evaluation. Popular ML models, e.g. ResNet18, ResNet50, ResNet152~\cite{he2016deep} and VGG16~\cite{simonyan2014very}, are trained with the dataset CIFAR-10~\cite{krizhevsky2009learning}, which has 60k images in total. In each set of experiments, we keep the random seed fixed and the learning rate initially set as 0.1 and multiplied by 0.8 for every 10 epochs to ensure the same variables. We present the results on ResNet50 and VGG16 in this section, and the other models show similar improvements.

\subsubsection{Performance Metrics}

Three metrics are used in the evaluation of LTP: 

\begin{itemize}
    \item Training throughput, which is the training speed \textit{(images/sec)} on DML training tasks.
    \item Time to accuracy (TTA), which is the time to reach the accuracy in the evaluation of test datasets.
    \item Batch synchronization time (BST), which is the sum of the times spent on gathering and broadcasting for each batch. It can reflect the efficiency of the communication protocol used during synchronization.
\end{itemize}

In addition to these three metrics, we evaluate the fairness when LTP and other commonly used transport protocols coexist.

\subsection{Training Throughput and Time to Accuracy}\label{sec:throughput_and_tta}

The main optimization of LTP is focused on transmission protocols. It is not uncommon to perform DML training tasks in networks where packet loss exists (\S~\ref{sec:tcp_poor_performance}). As a result, we compare LTP with commonly used congestion control algorithms in lossy networks with different packet loss rates (0\%, 0.01\%, 0.1\%, 0.5\%, 1\%). The choices of packet loss rates are also used in the evaluation of~\cite{lao2021atp}, and we will use them on the evaluations of BST as well (\S~\ref{sec:bst}).

\textbf{Improvement of Training Throughput.} Two models are used in the evaluation of throughput, ResNet50 (computation-intensive, with 98MB model size) and VGG16 (communication-intensive, with 500+MB model size). Results of throughput are shown in Figure~\ref{fig:throughput}, which demonstrates that LTP can achieve higher throughput in lossy networks. Compared with BBR in ResNet50, LTP achieves a 1.26x throughput improvement when the network has no non-congestion packet loss and better throughput improvement (up to 2x) when the non-congestion packet loss rate is from 0.01\% to 1\%. Cubic and New Reno are very sensitive to packet loss due to their congestion discovery mechanisms. Therefore, LTP can deliver performance improvements of up to 31x. The improvement of training throughput is not significant when using the VGG16 model. We think that this is due to
the model size of VGG16 being 5x greater than ResNet50. As a result, the communication size is 5x in the evaluation of VGG16. Elephant flows will reduce the hazard of long-tailed latency, which results in LTP having limited throughput improvement compared to BBR in the evaluation of VGG16. 

In our analysis, this performance improvement comes from 1) the introduction of the Early Close mechanism, which avoids the impact of long-tail latency on the BST; and 2) LTP's BDP-based congestion control mechanism, which can maintain a high bandwidth utilization in the network with non-congestion packet loss. 

\textbf{Precision Loss.} The introduction of bubble-filling can result in partial data loss during transmission. Therefore, we evaluate the impact of this partial data loss on the convergence of accuracy by the TTA. Figure~\ref{fig:time_to_accuracy} is the results of TTA between LTP and other congestion control algorithms, which show that LTP does not lead to a reduction in the top-1 accuracy within the typical range of packet loss rates. 

It is worth noting that random loss of gradients is a double-edged sword for DML training tasks. Each model has its unique tolerance for data loss, and when within the tolerance, partial gradient loss can enhance the generalization of the model, even allowing the model to reach a better accuracy eventually. Training convergence suffers when the number of gradients lost exceeds the model's tolerance, which varies between different models and needs to be discussed separately.

\subsection{Batch Synchronization Time}\label{sec:bst}

To analyze the fundamental reasons for the improvement in throughput brought by LTP, we collect the BST distribution during DML training and show them in the box plots (Figure~\ref{fig:batch_time}). The results show that the BST of LTP vastly outperforms the traditional TCP congestion control algorithms (Cubic and New Reno). Compared with BBR, the average BST of LTP is also reduced by about 30\%.

In the evaluation, the long-tail latency problem becomes more severe as the packet loss rate increases. This phenomenon does not occur in the evaluations of LTP because of the introduction of the Early Close mechanism, which can improve the FCT at the cost of partial data loss and significantly optimize the overall throughput of DML training.

\begin{figure}[tbp]
\centerline{\includegraphics[width=7cm]{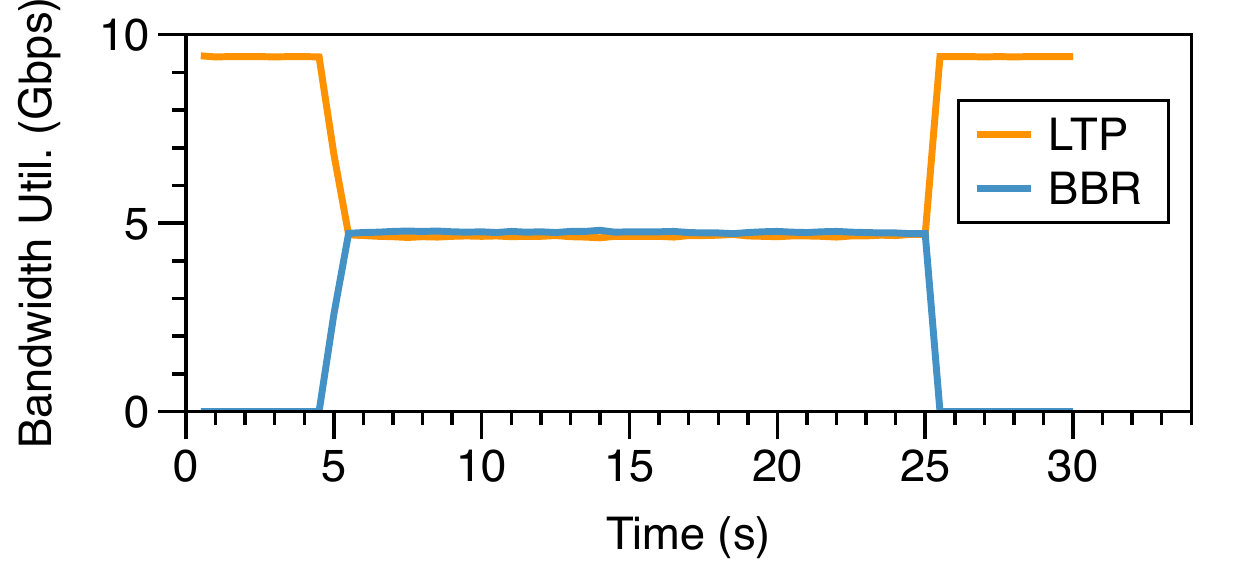}}
\caption{Fairness between BBR and LTP.}
\vspace{-0.1in}
\label{fig:fairness}
\end{figure}

\subsection{Fairness between Different Congestion Controls}\label{sec:fairness}

As a transport protocol with a customized congestion control algorithm, it is necessary to evaluate its fairness when coexisting with other protocols. We use point-to-point communication to evaluate the LTP's bandwidth allocation with other congestion control algorithms (including BBR, Cubic, and New Reno). The evaluation shows that LTP performs slightly worse than BBR in terms of bandwidth consumption (about 97\%, Figure~\ref{fig:fairness}). We speculate that the slight performance loss is due to the extra packet header (9B) in LTP. Further, due to the close performance of these two methods, LTP should exhibit a similar performance as BBR when coexisting with other congestion control protocols.  

\section{Discussion and Future Work}\label{sec:discussion}

\subsection{Collaboration with Other Gradient Compression Algorithms} \label{sec:coll_other_compression}

The approaches of gradient compression have similar ideas compared to LTP but have different optimization targets. The goal of LTP is to optimize the synchronization time of DML training on the network with multiple conditions, while the gradient compression approaches are to reduce the communication size of DML training. LTP can collaborate with the gradient compression works to further improve DML tasks' throughput. Such collaboration will introduce secondary compression, leading to further loss of gradients. One of our future works is to evaluate the performance gains achieved by these algorithms working together.

\subsection{Minimum Data Percentage on Different Models}

In the evaluations, when training the model ResNet50 under the network with a loss rate larger than 5\%, the problem of precision loss rarely happens. However, the same problems do not occur in the training tasks on the VGG16 model. We speculate that this is mainly due to the limited tolerance of data loss on the ResNet50 model, and the data discarded by the bubble-filling exceeds the tolerance, which in turn triggers problems in the final accuracy. Therefore, we are testing different models with different objectives to capture the approximate data loss tolerance range for each model, e.g. InceptionV3\cite{szegedy2016rethinking}, LSTM\cite{jozefowicz2016exploring} and NCF\cite{he2017neural}.

We are evaluating whether it is too aggressive to close the connection immediately when the transmission time reaches the deadline. One possible solution is still requiring a minimum arrival rate after the deadline is reached, rather than just ending the transmission. This minimum arrival rate may be related to the tolerance for data loss of the model, and we are conducting further experiments.

\subsection{LTP in DML Systems with Heterogeneous Networks.}

Network heterogeneity is common in federated learning and DML systems across DCs. LTP guarantees different data arrival rates under different quality networks by the LT Threshold. In heterogeneous network scenarios, different worker nodes may have different gradient arrival rates, which leads to bias in the contributions from different worker nodes. Network environments discussed in this paper are all homogeneous, so the arrival rates of gradients sent by worker nodes are relatively even. We are working on reducing the bias of the gradient contribution among different worker nodes on heterogeneous networks to improve the model's generalization.

\section{Related Works}\label{sec:related_works}

In recent years, DML has received extensive attention from industry and academia. However, the communication overhead between worker nodes and PSes has always been a severe bottleneck restricting the efficiency of DML training. Many optimization methods have been proposed.

\subsubsection{Gradient Compression}

Gradient compression is one of the methods that has been widely studied. This method can be divided into two main types. The first is gradient sparsification, which selects a portion of the data transferred from the gradient vector. The Top-k~\cite{aji2017sparse} algorithm transmits only the absolute values of the first k large in the gradient vector. Random-k~\cite{stich2018sparsified} randomly transmits a part of the data from the gradient vector, which reduces the sorting overhead compared to Top-k. Threshold-v~\cite{dutta2020discrepancy} transmits the absolute value of the gradient vector greater than a certain threshold, but this threshold is difficult to set appropriately. The second method is gradient quantization. 
8-bit quantization~\cite{dettmers20158} converts each 32-bit floating number to 8-bit. Seide~\cite{seide20141} et al. propose an extreme form of quantization: convert all gradient values to 0 and 1 according to the range of gradient elements. However, these methods all have the potential to introduce a loss of accuracy, and the error compensation technique~\cite{tang2019doublesqueeze} proved to be an effective method for this problem. Gradient compression is fully analyzed and compared in detail in GRACE~\cite{xu2021grace}.

As mentioned at \S~\ref{sec:coll_other_compression}, LTP can collaborate with these approaches to further reduce synchronization overhead. We are evaluating whether the use of LTP on these approaches results in a precision loss.

\subsubsection{Network Acceleration of DML}

Another effective way to improve synchronization efficiency is to optimize or adjust the transmission network. Xia \textit{et al.}~\cite{xia2019rethinking} verify that DML tasks can tolerate partial packet loss and call on the community to develop a transmission protocol that supports packet loss tolerance. Based on the same finding, DGT~\cite{zhou2021dgt} divides gradients into two categories based on importance. Necessary gradients are transmitted over reliable TCP, while unimportant gradients are transmitted over unreliable UDP. It can improve the throughput of DML training, but it lacks discussions on arriving thresholds and has no detailed design of congestion control for the UDP channel. SwitchML~\cite{sapio2019scaling} and ATP~\cite{lao2021atp} use programmable switches to achieve in-network aggregation, which significantly improves the efficiency of model training.

\section{Conclusion}\label{sec:conclusion}

We design a transmission protocol named LTP to accelerate the DML training tasks with PS architecture. The target of LTP is to reduce the impact of incast traffic patterns generated by the PS architecture's communication pattern. By enabling the loss-tolerant transmission, LTP allows partial loss of synchronization gradients during DML training, leading to faster synchronization in each iteration. LTP uses the Early Close mechanism to tune the threshold of loss-tolerant transmission and utilizes the bubble-filling for lost data correction. A prototype of LTP is implemented and integrated into the widely used ML framework PyTorch. Real testbed evaluations with 9 machines on popular ML models show that LTP can achieve up to 30x speedup on DML training throughput compared to traditional TCP congestion control algorithms and can achieve up to 2x speedup compared to BBR without accuracy loss.



\ifCLASSOPTIONcaptionsoff
  \newpage
\fi

\bibliographystyle{Scripts/Touko-Format-unsrt-abbrv}
\bibliography{main}

\end{document}